\documentclass[aps,prc,preprint,eqsecnum,showpacs,floatfix]{revtex4-1}

\usepackage{graphicx,epstopdf}
\usepackage{hyperref}
\usepackage{color}
\usepackage{subfigure}
\usepackage{amsmath}
\usepackage{amssymb}
\usepackage{epsfig}
\usepackage{amsfonts}
\usepackage{mathtools}
\usepackage{bbm}
\usepackage{bm}
\usepackage{float}
\usepackage{xcolor}

\def\rvec{{\bf r}}
\def\hvec{{\bf h}}
\def\kvec{{\bf k}}
\def\pvec{{\bf p}}
\def\qvec{{\bf q}}
\def\Lvec{{\bf L}}
\def\bra#1{\left\langle#1\right|}
\def\ket#1{\left|#1\right\rangle}

\def\he#1{$^{#1}$He}

\def\Im{{\cal I}m}
\def\I{{\rm i}}

\def\Tr{{\cal T}r}
\def\KF{k_{\rm F}}
\def\SF{S_{\rm F}}
\def\tF{t_{\rm F}}
\def\1{\mathbbm{1}}

\def\bra#1{\bigl\langle{ #1} \bigr|}
\def\ket#1{\bigl|{ #1} \bigr\rangle}

\def\ovlp#1#2{\bigl\langle{ #1}\big|{#2} \bigr\rangle}

\def\etal{{\em et al.\/}\ }
\def\ie{{\em i.e.\/}\ }
\def\cf{{\em cf.\/}\ }

\def\sij#1#2{{\bm\sigma}_{#1}\cdot{\bm\sigma}_{#2}}

\def\boxit#1{
        \centerline{\vbox{\hsize=6.0truein\hrule\hbox{\vrule\kern5pt
        \vbox{\kern5pt\noindent #1\smallskip
        \kern5pt}\kern5pt\vrule}\hrule}
}}
\def\he#1{$^{#1}$He}

\begin{document}

\title{Variational and parquet-diagram calculations for neutron
  matter.\\ I.  Structure and Energetics.}

\author{E.~Krotscheck and J. Wang}

\affiliation{Department of Physics, University at Buffalo, SUNY
Buffalo NY 14260}
\affiliation{Institut f\"ur Theoretische Physik, Johannes
Kepler Universit\"at, A 4040 Linz, Austria}

\begin{abstract}
  We develop the variational/parquet diagram approach to the structure
  of nuclear systems with strongly state-dependent interactions. For
  that purpose, we combine ideas of the general Jastrow-Feenberg
  variational method and the local parquet-diagram theory for bosons
  with state-dependent interactions (R. A. Smith and A. D. Jackson,
  Nucl. Phys. {\bf 476}, 448 (1988)). The most tedious aspect of
  variational approaches, namely the symmetrization of an operator
  dependent variational wave function, is thereby avoided.
  We carry out calculations for neutron matter interacting via the
  Reid and Argonne $v_6$ models of the nucleon-nucleon interaction.
  While the equation of state is a rather robust quantity that comes
  out reasonably well even in very simplistic approaches, we show that
  effective interactions, which are the essential input for
  calculating dynamic properties, depend sensitively on the quality of
  the treatment of the many-body problem.
\end{abstract}

\maketitle

\section{Introduction}
Realistic nucleon-nucleon interactions depend on the
relative spin, isospin, orientations, and angular momenta of the
nucleons involved. Popular phenomenological models
\cite{Reid68,Bethe74,Day81,AV18,Wiri84} represent the interaction in the form
of a sum of local functions, times correlation operators, \ie
\begin{equation}
\hat v (i,j) = \sum_{\alpha=1}^n v_\alpha(r_{ij})\,
        \hat O_\alpha(i,j),
\label{eq:vo}
\end{equation}
where $r_{ij}=|\rvec_i-\rvec_j|$ is the distance between particles $i$
and $j$, and the $O_\alpha(i,j)$ are operators acting on the spin,
isospin, and possibly angular momentum variables of the individual
particles.  According to the number of operators $n$, the potential
model is referred to as a $v_n$ model potential. Semi-realistic models
for nuclear matter keep at least 6 operators, but up to 28 operators
have been included \cite{Wiri84} in the sum (\ref{eq:vo}).

The six base operators are
\begin{eqnarray}
O_1(i,j;\hat\rvec_{ij})
        &\equiv& O_c = \1,
\nonumber\\
O_3(i,j;\hat\rvec_{ij})
        &\equiv& ({\bm\sigma}_i \cdot {\bm\sigma}_j)\,,
\nonumber\\
O_5(i,j;\hat\rvec_{ij})
&\equiv& S(i,j;\hat\rvec_{ij})
      \equiv 3({\bm\sigma}_i\cdot \hat\rvec_{ij})
      ({\bm\sigma}_j\cdot \hat\rvec_{ij})-{\bm\sigma}_i \cdot {\bm\sigma}_j\,,
      \nonumber\\
      O_{2n}(i,j;\hat\rvec_{ij}) &=& O_{2n-1}(i,j;\hat\rvec_{ij})
      {\bm\tau}_1\cdot{\bm\tau}_2\,.
\label{eq:operator_v6}
\end{eqnarray}
where $\hat\rvec_{ij} = \rvec_{ij}/r_{ij}$.  These operators
are referred to as central, spin, tensor, isospin, spin-isospin and
tensor-isospin operators, respectively.  The arguments $i,j$ and
$\hat\rvec_{ij}$ of state-dependent functions will be omitted for
simplicity when no ambiguity arises.

For simple, state-independent interactions as appropriate for electrons
or quantum fluids, the Jastrow-Feenberg ansatz \cite{FeenbergBook}
for the wave function
\begin{equation}
\Psi_0 = \prod^N_{i,j=1 \atop i<j} f(r_{ij})\Phi_0
\label{eq:Jastrow}
\end{equation}
and its logical generalization to multiparticle correlation functions
has been extremely successful.  Here $\Phi_0$ is a model state
describing the statistics and, when appropriate, the geometry of the
system; for fermions it is normally taken as a Slater determinant.
One of the reasons for the success of this wave function is that it
provides a reasonable upper bound for the ground state energy
\begin{equation}
E_0 = \frac{\bra{\Psi_0}H\ket{\Psi_0}}{\ovlp{\Psi_0}{\Psi_0}}\,.
  \label{eq:energy}
\end{equation}
It has therefore been applied in both semi-analytic calculations
\cite{FeenbergBook} as well as early Monte Carlo calculations
\cite{KalosLevVer,CeperleyVMC} and is still being used as an importance
sampling function for diffusion and Green's Functions Monte Carlo
computations \cite{CeperleyRMP,JordiQFSBook}. Semi-analytic methods
employ diagram expansions and integral equation methods -- specifically
the hypernetted chain (HNC) summations \cite{Morita58,LGB} or its
their fermion versions (FHNC) \cite{Mistig,Fantoni} -- for the calculation of
physically interesting quantities. 

In particular, this approach permits an unconstrained optimization of 
the assumed correlation functions,
\begin{equation}
\frac{\delta E_0}{\delta f_n}({\bf r}_1,\ldots,{\bf r}_n) = 0,
\label{eq:euler}
\end{equation}
in which case the method is referred to as the (Fermi-)Hypernetted-Chain-Euler-Lagrange (F)HNC-EL procedure.  It has been a particularly important 
insight that the HNC-EL method corresponds, for bosons, to a self-consistent 
summation of all ring and ladder diagrams of perturbation theory --
the so-called ``parquet'' diagrams \cite{parquet1,parquet2,parquet3}. 
To carry out these summations, specific local approximations are made, 
but the upper-bound property for the energy makes sure that one has 
achieved the best approximation for the computational price one is 
willing to pay.

The Jastrow-Feenberg ansatz (\ref{eq:Jastrow}) is insufficient for
dealing with realistic nucleon-nucleon interactions of the form
(\ref{eq:vo}).  A logical generalization of the Jastrow-Feenberg wave
function (\ref{eq:Jastrow}) is the symmetrized operator product
\cite{FantoniSpins,IndianSpins}
\begin{equation}
        \Psi_0^{{\rm SOP}}
        = {\cal S} \Bigl[ \prod^N_{i,j=1 \atop i<j} \hat f (i,j)\Bigr] \Phi_0\,,
\label{eq:f_prodwave}
\end{equation}
where
\begin{equation}
  \hat f(i,j) = \sum_{\alpha=1}^n f_\alpha(r_{ij})\,
  O_\alpha(i,j)
\end{equation}
and ${\cal S}$ stands for symmetrization. Unfortunately, the need to
symmetrize the correlations figuratively opens Pandora's box. Not
surprisingly, only limited success has been achieved
\cite{FantoniSpins,Wiri78,IndianSpins}. In fact, it is not even clear
how to choose the correlation functions $f_\alpha(r_{ij})$ because,
due to the symmetrization, components $v_\alpha(r)$ of the interaction
are multiplied, in the energy expectation value (\ref{eq:energy}), by
products of correlation functions $f_\beta(r) f_\gamma(r)$ with
$\beta\ne\alpha$ and $\gamma\ne\alpha$ \cite{SpinTwist}.  This makes
the use of simplistic choices of the correlation functions like the
``low-order constrained variational (LOCV) method''
\cite{ScottMozowski,PandharipandeBethe} highly problematic if the
interactions in the different operator channels are very different
\cite{SpinTwist} and sufficiently high-order commutators are
included. Hence, only very simple implementations -- the so-called
``single operator chain (SOC)'' approximation \cite{PAW79} -- have
been carried out.  Moreover, the complicated structure of commutator
terms makes the identification with Feynman diagrams nontransparent.

In view of this situation, Smith and Jackson \cite{SmithSpin} started
from the idea of localized parquet-diagram summations and implemented
the procedure for a fictive system of bosonic nucleons interacting via
a $v_6$ interaction. It turned out that the equations derived were
identical to those one would obtain in a bosonic version of the
summation method of Fantoni and Rosati \cite{FantoniSpins}, which
simply ignored the fact that the individual pair correlation operators
$\hat f(i,j)$ do not commute. In other words, the problem of the
importance of commutator diagrams does not go away, but the idea of
parquet-diagram summations promises a clearer procedure to deal with
these effects without having to go through the development of a full
variational procedure.

Therefore we adopt here the ideas of Smith and Jackson and generalize
them to Fermi systems. To that end, in the next section we will first
review how the connections between the HNC-EL equations of the
Jastrow-Feenberg theory and parquet-diagram summations are
established.  We will then show how specific equations from the
FHNC-EL theory for state-independent correlations can be derived from
corresponding sets of parquet diagrams.  Specifically, we will focus
on ring diagrams and the Bethe-Goldstone equation. Once the
connections have been established, we can go on and formulate the
method for fermions with a $v_6$ interaction.

We will restrict ourselves to neutron matter in our applications for
two reasons: First, we feel that the problem of commutator diagrams
which would, in the language of parquet theory, correspond to
``twisted'' ladder rungs, is solved.  Additionally, it has been
pointed out \cite{Baldo2012} that the spin-orbit force, which is
omitted in the $v_6$ models, plays an important role in nuclear matter
calculations near isospin symmetry.  We will demonstrate the
importance of both the state-dependence of the correlations and the
non-locality introduced by the Pauli principle.  On the other hand, we
do not attempt an exhaustive comparison with previous neutron-matter
calculations as carried out in Ref.~\onlinecite{Baldo2012}. Rather, we
concentrate on the technical implementations of parquet theory and its
connections to FHNC.

\section{Variational and parquet theory}
\subsection{Bosons in a nutshell}
In this subsection we review very briefly the optimized variational
method (``HNC-EL'') for bosons and the ``local parquet summations,''
because these equations have a very familiar structure and can be
derived with minimal approximations from various basic
theories. Skipping the technical details that can be found in original
papers and pedagogical expositions, we display the resulting
equations.

The static structure function $S(q)$ is expressed in terms of a Bogoliubov 
equation
\begin{equation}
  S(q) = \frac{1}{\sqrt{1+\displaystyle\frac{2\tilde V_{\rm p-h}(q)}{t(q)}}}
  \label{eq:BoseRPA}
\end{equation}
in terms of a self-consistently determined ``particle-hole''
interaction $V_{\rm p-h}$.  Having defined a dimensionless Fourier 
transform by including a density factor $\rho$, i.e.,
\begin{equation}
  \tilde f(q) = \rho\int d^3r e^{\I \qvec\cdot\rvec}f(r)\label{eq:ft}\,,
\end{equation}
this effective interaction takes the specific form
\begin{eqnarray}
  V_{\rm p-h}(r) &=& g(r)\left[v(r) + \Delta V_e(r)\right]
  + \frac{\hbar^2}{m}\left|\nabla\sqrt{g(r)}\right|^2\nonumber\\
  && + \left[g(r)-1\right]w_{\rm I}(r)\,,\label{eq:BoseVph}\\
  \tilde w_{\rm I}(q) &=& -t(q)
\left[1-\frac{1}{ S(q)}\right]^2
\left[S(q)+\frac{1}{2}\right]\,.
  \label{eq:Bosewind}
\end{eqnarray}
In the language of Jastrow-Feenberg theory, the term $\Delta V_e(r)$ 
accounts for the contribution from ``elementary diagrams'' and multiparticle
correlations \cite{EKthree}, whereas in terms of parquet-diagram 
theory it is the contribution of diagrams that are both particle-particle 
and particle-hole irreducible \cite{TripletParquet}.

A few algebraic manipulations show that the pair distribution function
satisfies the coordinate-space equation \cite{LanttoSiemens}
\begin{equation}
  \frac{\hbar^2}{m}\nabla^2\sqrt{g(r)} = \left[v(r) + \Delta V_e(r) +
    w_{\rm I}(r)\right]\sqrt{g(r)}
  \label{eq:BoseBG}\,.
  \end{equation}
Eq.~(\ref{eq:BoseBG}) is recognized as the boson Bethe-Goldstone
equation in terms of the interaction $v(r) + \Delta V_e(r) + w_{\rm I}(r)$. 
This observation led Sim, Woo, and Buchler \cite{Woo70} to
the conclusion that ``it appears that the optimized Jastrow function
is capable of summing all rings and ladders, and partially all other
diagrams, to infinite order.'' In fact, the form of the equations
can be obtained by {\em demanding\/} that the pair distribution
function $g(r)$ satisfies both the Bogoliubov equation (\ref{eq:BoseRPA})
and the Bethe-Goldstone equation (\ref{eq:BoseBG}) \cite{PairDFT};
the only quantity undetermined by that requirement is $\Delta V_e(r)$.

Since we will heavily rely on the derivations and localization procedures 
of parquet-diagram theory, let us briefly review the relevant steps.  
First, the Bogoliubov equation (\ref{eq:BoseRPA}) is derived from a 
random-phase approximation (RPA) equation for the
density-density response function
 \begin{eqnarray}
    \chi(q,\omega) &=&
    \frac{\chi_0(q,\omega)} {1-\tilde V_{\rm
        p-h}(q)\chi_0(q,\omega)}\label{eq:chiRPA}\,\\
 S(q) &=& -\int_0^\infty \frac{d\hbar\omega}{\pi} \Im \chi(q,\omega),
\label{eq:SRPA}
 \end{eqnarray}
in terms of a {\em local\/} and {\em energy-independent\/} particle-hole
interaction $\tilde V_{\rm p-h}(q)$. Here
\begin{equation}
        \chi_0(q,\omega) =
        \displaystyle \frac{2 t(q)}
        { (\hbar\omega+\I\eta)^2-
        t^2(q)} \,,
\label{eq:Chi0Bose}
\end{equation}
with $t(q) = \frac{\hbar^2 q^2}{2m}$, is the particle-hole propagator
of non-interacting bosons. Eq.~(\ref{eq:SRPA}) defines an {\em energy
  dependent\/} effective interaction
 \begin{equation}
   \widetilde W(q,\omega) = \frac{\tilde V_{\rm p-h}(q)} {1-\tilde V_{\rm
       p-h}(q)\chi_0(q,\omega)}\,.
   \label{eq:Wnonlocal}
 \end{equation}
An {\em energy independent\/} effective interaction $\widetilde W(q)$ is then
defined such that it leads to the same $S(q)$, {\ie \/}
\begin{eqnarray}
S(q) &=& -\int_0^\infty \frac{d\hbar\omega}{\pi} \Im
 \frac{\chi_0(q,\omega)} {1-\tilde V_{\rm
     p-h}(q)\chi_0(q,\omega)}\nonumber\\
 &=& -\int_0^\infty \frac{d\hbar\omega}{\pi} \Im
 \left[\chi_0(q,\omega)+ \chi^2_0(q,\omega) \widetilde W(q,\omega)\right]
 \nonumber\\
  &\overset{!}{=}& 
-\int_0^\infty \frac{d\hbar\omega}{\pi} \Im
\left[\chi_0(q,\omega)+ \chi^2_0(q,\omega) \widetilde W(q)\right]\,,
\label{eq:Wlocal}
\end{eqnarray}
where the last line defines $\widetilde W(q)$. Carrying out the
integration leads to
\begin{equation}
  \widetilde W(q) = -t(q)(S(q)-1)\label{eq:BoseWind}\,.
\end{equation}
The particle-hole reducible part 
\begin{equation}
  \tilde w_I(q) = \widetilde W(q) - \tilde V_{\rm p-h}(q)
\label{eq:wind}
\end{equation}
of $\widetilde W(q)$ so defined is then exactly the induced
interaction (\ref{eq:Bosewind}).  This local $w_I(r)$ then supplements
the bare interaction in the Bethe-Goldstone equation.

It is relatively straightforward to generalize the procedure to
interactions with spin and tensor components that are needed for
nuclear systems \cite{SmithSpin}.
  
\subsection{Fermions with state-independent interactions}

We discuss here the simplest implementation of the FHNC theory that is
compatible with the variational problem, called the FHNC//0 approximation.
This version has quantitative deficiencies, in particular at high densities, 
but it permits the clearest connection to the summation
of ring and ladder diagrams of parquet theory.  The implementation
and relevance of higher order exchange corrections will be discussed
below in Section \ref{ssec:exchanges}.

In the FHNC//0 approximation, the generalization of Eq.~(\ref{eq:BoseRPA})
is
\begin{equation}
  S(q) = \frac{\SF(q)}{\sqrt{1 + 2 \displaystyle\frac{\SF^2(q)}{t(q)}
      \tilde  V_{\rm p-h}(q)}}\,.
\label{eq:FermiPPA}
\end{equation}
where
\begin{equation}
  \SF(q) = \begin{cases}
     \displaystyle \frac{3q}{4\KF}-\frac{q^3}{16\KF^3},
                & q < 2\KF ;\\
      1,      & q \ge 2\KF\,.
      \end{cases}
\end{equation}
is the static structure function of the non-interacting Fermi gas.

In this approximation, the effective interaction $\tilde V_{\rm
  p-h}(q)$ is approximated by the ``direct'' part of the particle-hole
interaction, $\tilde V_{\rm p-h}(q)\approx \tilde V_{\rm dd}(q)$ in
the language of the FHNC summations \cite{Johnreview,polish}. This
quantity is structurally identical to that for bosons, \ie,
\begin{equation}
  V_{\rm p-h}(r) = V_{\rm CW}(r)+
  \Gamma_{\!\rm dd}(r)w_{\rm I}(r)\label{eq:Vph}\,.
\end{equation}
Here
\begin{equation}
V_{\rm CW}(r) = (1+\Gamma_{\!\rm dd}(r))v(r)+
\frac{\hbar^2}{m}\left|\nabla\sqrt{1+\Gamma_{\!\rm dd}(r)}\right|^2
\label{eq:vcw}
\end{equation}
is the ``Clark-Westhaus'' effective interaction \cite{Johnreview},
$\Gamma_{\!\rm dd}(r)$ is the so-called direct correlation function,
and
$w_{\rm I}(r)$ is the ``induced interaction''
\begin{eqnarray}
  \tilde w_{\rm I}(q) &=&-V_{\rm p-h}(q) - t_{\rm F}(q)
  \tilde\Gamma_{\!\rm dd}(q)
  \nonumber\\
  &=& -t(q)
\left[\frac{1}{\SF(q)}-\frac{1}{ S(q)}\right]^2
\left[\frac{S(q)}{\SF(q)}+\frac{1}{2}\right]\,,
  \label{eq:Fermiwind}
\end{eqnarray}
where we have abbreviated
\begin{equation}
  \tF(q) = \frac{t(q)}{\SF(q)} \label{eq:tFdef}
\end{equation}
for future reference.  In FHNC//0 approximation, the static structure 
function $S(q)$ and the direct correlation function 
$\tilde\Gamma_{\!\rm dd}(q)$ are related by
\begin{equation}
  S(q) = \SF(q)\left[1+\tilde \Gamma_{\!\rm dd}(q)\SF(q)\right]\,.
  \label{eq:SFHNC0}
\end{equation}
The Bose limit is obtained by setting $\SF(q) \rightarrow 1$. Note
that the Fourier transform of Eq.~(\ref{eq:SFHNC0}) does not give a
useful expression for the pair distribution function. This feature,
its cause, and how to overcome it has been discussed in many places;
see, for example, Refs.~\onlinecite{polish} and \onlinecite{fullbcs}.
 
To derive the equation determining the short-ranged structure of the
correlations, we begin with Eq.~(\ref{eq:Fermiwind}) which, using
Eqs.~(\ref{eq:Vph}) and (\ref{eq:vcw}), can be written as
\begin{eqnarray}
V_{\rm p-h}(r) +   w_{\rm I}(r) &=&
(1+\Gamma_{\!\rm dd}(r))\left[v(r)+w_{\rm I}(r)\right]\nonumber\\
&+&
\frac{\hbar^2}{m}\left|\nabla\sqrt{1+\Gamma_{\!\rm dd}(r)}\right|^2\nonumber\\
&=& -\left[\tF(q)\tilde\Gamma_{\!\rm dd}(q)\right]^{\cal F}(r)
  \label{eq:windBG}
\end{eqnarray}
(\cf Eq.~(2.62)) of Ref.~\citenum{mixmonster}).
This expression can in turn be rewritten in coordinate space as
\begin{eqnarray}
&&\sqrt{1+\Gamma_{\!\rm dd}(r)}\left[-\frac{\hbar^2}{2m}\nabla^2 + v(r) + w_{\rm I}(r)\right]
  \sqrt{1+\Gamma_{\!\rm dd}(r)}\nonumber\\
  &=&\left[t_F(q)(\SF(q)-1)\tilde\Gamma_{\!\rm dd}(q)\right]^{\cal F}(r)\,,
  \label{eq:ELSchr}
\end{eqnarray}
where $\left[\dots\right]^{\cal F}$ stands for the Fourier transform
(\ref{eq:ft}).

The right-hand side of Eq.~(\ref{eq:ELSchr}) is evidently zero for bosons, 
and the Euler equation is a simple zero-energy Schr\"odinger equation 
where the bare interaction is supplemented by the induced potential. 
For fermions, the right-hand side alters \cite{OwenVar} the short-ranged 
behavior of the correlation function $\Gamma_{\!\rm dd}(r)$, and hence the
short-ranged behavior of the pair distribution function $g(r)$.

\subsection{Connections between FHNC and parquet diagrams}
\label{sec:parquet}

\subsubsection{Rings}
\label{ssec:rings}

The expression (\ref{eq:FermiPPA}) reduces to the Bogoliubov equation
for the case of bosons, with $\SF(q)=1$. For fermions, we must identify
$\chi_0(q,\omega)$ with the Lindhard function
\begin{equation}
  \chi_0(q,\omega)
  =\frac{2}{N}\sum_{\hvec}
  \frac{n(h)\bar n(|\hvec+\qvec|)(t(|\hvec+\qvec|)-t(h))}{
    (\hbar\omega+\I\eta)^2 - (t(|\hvec+\qvec|)-t(h))^2}\,,
  \label{eq:Lindhard}
\end{equation}
where $n(q) = \theta(\KF-q)$ is the Fermi distribution and $\bar
n(q) = 1-n(q)$.  Consistent with the convention (\ref{eq:ft})
(for which $\tilde V_{\rm p-h}(q)$ has the dimension of an
energy), we have defined the density-density response function slightly
differently than usual \cite{FetterWalecka}, namely such that it has the
dimension of an inverse energy.

The energy integration can no longer be carried out analytically.
Nevertheless, we anticipate that Eq.~(\ref{eq:FermiPPA}) can also be
derived for fermions from the random-phase approximation
(\ref{eq:SRPA}) for the dynamic structure function.  Given any
function $f(\pvec,\hvec)$ depending on a ``hole momentum''
$|\hvec|<\KF$ and a ``particle momentum'' $\pvec=\hvec+\qvec$ with
$|\pvec|>\KF$, we may define its Fermi-sea average by
\begin{eqnarray}
\left\langle f(\pvec,\hvec) \right\rangle(q)
&=& \frac{\sum_{\hvec} \bar n(|\hvec+\qvec|) n(h) f(\hvec+\qvec,\hvec)}
{\sum_{\hvec} \bar n(|\hvec+\qvec|) n(h)}\label{eq:favg}\\
&=& \frac{1}{S_{\rm F}(q)}\int \displaystyle\frac{d^3h}{V_{\rm F}}
\bar n(|\hvec+\qvec|) n(h)  f(\hvec+\qvec,\hvec)\,,\nonumber
\end{eqnarray}
where $V_{\rm F}$ is the volume of the Fermi sphere.
In particular, we find
\begin{equation}
\left\langle t(|\hvec+\qvec|)-t(h)\right\rangle(q)
= \frac{t(q)}{\SF(q)} = \tF(q)\label{eq:tqcoll}
\end{equation}
which justifies our identification of $\tF(q)$ as an ``average''
kinetic energy of the non-interacting Fermi system.

Eq.~(\ref{eq:FermiPPA}) can then be obtained by approximating the
particle-hole energies $t(|\hvec+\qvec|)-t(h)$ in the Lindhard
function (\ref{eq:Lindhard}) by the ``average'' kinetic energy
$\tF(q)$, leading to a ``collective'' Lindhard function,
\begin{equation}
        \chi_0^{\rm coll}(q,\omega) =
        \displaystyle \frac{2 t(q)}
        { (\hbar\omega+\I\eta)^2-
        \tF^2(q)} \,.
\label{eq:Chi0Coll}
\end{equation}
This approximation is occasionally also referred to as a ``one-pole
approximation'' or ``mean spherical approximation'.  Alternative
rationalizations of the collective approximation for the Lindhard
function may be found in Ref.~\onlinecite{fullbcs}.  The frequency
integration (\ref{eq:SRPA}) can then be carried out analytically and
leads to equation (\ref{eq:FermiPPA}).
 
\subsubsection{Ladder Rungs}
\label{ssec:rungs}

The analysis leading to the identification of a local,
energy-independent induced interaction $\tilde w_{\rm I}(q)$ based on
an energy-dependent interaction of the form (\ref{eq:Wnonlocal}) is
exactly the same for fermions and for bosons.  Following
Refs.~\citenum{parquet1,parquet2}, we define a {\em local\/} effective
interaction through the condition (\ref{eq:Wlocal}),
\begin{eqnarray}
S(q) &=&\SF(q) - \widetilde W(q)\int_0^\infty \frac{d\hbar\omega}{\pi}\,
\Im  \chi_0^2(q,\omega) \,.
\label{eq:Scond}
 \end{eqnarray}
For further reference, let
\begin{equation}
  \int_0^\infty \frac{d\hbar\omega}{\pi}\Im  \chi_0^2(q,\omega)
  \equiv \frac{\SF^3(q)}{t(q)\lambda(q)}\,;
   \label{eq:lambdadef}
\end{equation}
we then have
\begin{equation}
\widetilde W(q) = -\tF(q)\lambda(q)\frac{S(q)-\SF(q)}{\SF^2(q)}\,.
  \label{eq:Wfermi}
\end{equation}
The frequency integral in Eq.~(\ref{eq:lambdadef}) can be carried out
analytically \cite{FetterWalecka}.  In the ``collective approximation'' 
(\ref{eq:Chi0Coll}) for $\chi_0(q,\omega)$, we obtain
$\lambda(q)=1$ and we recover the induced interaction $\tilde w_{\rm
  I}(q)$ from Eq.~(\ref{eq:Fermiwind}).

An issue that needs to be addressed when moving from the
Jastrow-Feenberg description to parquet diagrams concerns the definition of
$\tilde\Gamma_{\!\rm dd}(q)$. In FHNC//0 we can obtain this quantity
from $S(q)$ via Eq.~(\ref{eq:SFHNC0}).
To construct the equivalent of this relationship in parquet theory, we
go back to Eq.~(\ref{eq:Scond}). There we should identify
\begin{equation}
  \tilde\Gamma_{\!\rm dd}(q)\SF^2(q) = -\widetilde W(q)
  \int_0^\infty \frac{d\hbar\omega}{\pi} \Im\chi_0^2(q,\omega)
  = S(q)-\SF(q)\,.
\end{equation}
Accordingly, the relationship between $\tilde\Gamma_{\!\rm dd}(q)$ and $S(q)$ is
always given by Eq.~(\ref{eq:SFHNC0}).

\subsubsection{Ladders}
\label{ssec:ladders}

The final objective is to identify the coordinate-space equation with
a local approximation of the Bethe-Goldstone equation, whose exact
form still needs to be determined.  We begin with the Bethe-Goldstone
equation as formulated in Eqs.~(2.1), (2.2) of
Ref.~\citenum{BetheGoldstone57}. It is understood that $\pvec, \pvec'$
are particle states and $\hvec, \hvec'$ are hole states. Vectors
$\kvec,\ \kvec'$ can be either particle or hole states. Following
Ref.~\citenum{BetheGoldstone57}, we introduce the pair wave function
$\psi$ in a coordinate frame centered at the origin of the Fermi sea,
given by the integral equation
\begin{eqnarray}
  \bra{\kvec,\kvec'}\psi\ket{\hvec,\hvec'}
  &=&   \ovlp{\kvec,\kvec'}{\hvec,\hvec'}\label{eq:fullpsi2}\\
  &-& \bar n(k)\bar n(k')\frac{\bra{\kvec,\kvec'}v\psi
    \ket{\hvec,\hvec'}}{t(k) + t(k') -t(h)-t(h')}
\nonumber\,.
\end{eqnarray}

In making the connection to FHNC-EL, we should assume that the pair
wave function is a function of the relative coordinate
(or momentum), \ie
\[
  \bra{\kvec,\kvec'}\psi\ket{\hvec,\hvec'} = \frac{1}{N}\tilde\psi(q)
  \]
and set
\begin{equation}
  \psi(r) = \sqrt{1+\Gamma_{\!\rm dd}(r)}\,.
\end{equation}
Similarly, for local interactions, we should have
\[
\bra{\kvec,\kvec'}v\psi\ket{\hvec,\hvec'} = \frac{1}{N}
\left[v(r)\psi(r)\right]^{\cal F}(q)\,.
\]
To ensure this, the energy denominator coefficient
\[ \bar n(k)\bar n(k')\frac{\bra{\kvec,\kvec'}v\psi
    \ket{\hvec,\hvec'}}{t(k) + t(k') -t(h)-t(h')}
  \]
must somehow be approximated by a function of momentum transfer
$q$. This can be achieved by the averaging procedure (\ref{eq:favg}) 
applied to the above energy denominator coefficient, to yield 
\begin{equation}
2 \tF(q)\lambda(q)\left[\psi(q)-\delta(q)\right] = -[v(r)\psi(r)]^{\cal F}(q)\,.
  \label{eq:BGSchq}
\end{equation}
Alternatively, and more in
the spirit of Bethe and Goldstone, we write Eq.~(\ref{eq:fullpsi2}) as
\begin{eqnarray}
   &&\left[t(k) + t(k') -t(h)-t(h')\right]
\left[\bra{\kvec,\kvec'}\psi\ket{\hvec,\hvec'}-
\ovlp{\kvec,\kvec'}{\hvec,\hvec'}\right]\nonumber\\
  &&=-\bar n(k)\bar n(k')\bra{\kvec,\kvec'}v\psi\ket{\hvec,\hvec'}
  \label{eq:BGpsi3}\,.
\end{eqnarray}
The approximation
\begin{equation}
  t(|\hvec+\qvec|) -t(h)
  \approx \left\langle t(|\hvec+\qvec|) -t(h)
  \right\rangle(q) = \tF(q)\,.
\end{equation}
now gives Eq.~(\ref{eq:BGSchq}) without the factor $\lambda(q)$
or, in coordinate space, we have
\begin{eqnarray}
  &&\left[-\frac{\hbar^2}{m}\nabla^2 + v(r)\right]\psi(r)\nonumber\\
  &&= \left[2\tF(q)(\SF(q)-1)(\tilde\psi(q)-\delta(q))\right]^{\cal F}(r)\,.
  \label{eq:BGSchr}
\end{eqnarray}

Eq.~(\ref{eq:BGSchr}) is similar to, but not identical with, 
(\ref{eq:ELSchr}), which is obtained by the further assumption 
$\psi^2(r)-1 \ll 1$.  It therefore makes sense to assert $\psi(r) \approx
\sqrt{1+\Gamma_{\!\rm dd}(r)}$.  More importantly, the bare interaction
of the Bethe-Goldstone equation is supplemented by the induced
interaction.  The identification between the two expressions
(\ref{eq:BGSchr}) and (\ref{eq:ELSchr}) is not as precise as in the
case of the ring diagrams, but note that FHNC-EL//0 contains more than
just particle-particle ladders, also including particle-hole and hole-hole
ladders \cite{Rip79}.

\subsection{Propagator corrections}

Our analysis has so far addressed the question ``what does it take to
obtain a specific set of FHNC-EL diagrams from a corresponding set of
Feynman diagrams ?'' The analysis can be carried farther to other sets
of diagrams.  For example, the ``cyclic chain'' diagrams of the FHNC-EL
method can be derived from the self-energy diagrams by the same 
localization procedure described above. 

Once the approximations have been identified, it is also clear how to
improve upon them: There is no reason to use the ``collective
approximation'' (\ref{eq:Chi0Coll}) in both the frequency integrals
(\ref{eq:SRPA}) and the definition of the local effective interaction
(\ref{eq:Wlocal}). 

A second issue is then the generalization of the kinetic energy term
$\left|\nabla\sqrt{1+\Gamma_{\!\rm dd}(\rvec)}\right|^2$. To
this end, we begin with the localized Bethe-Goldstone equation
(\ref{eq:BGSchq}), where we supplement the bare interaction $v(r)$ by
the induced interaction $w_{\rm I}(r)$ and identify the pair wave
function $\psi(r)$ with $\sqrt{1+\Gamma_{\!\rm dd}(\rvec)}$. We then
have two equations, namely (\ref{eq:Wfermi}), which can be written as
\begin{equation}
  \widetilde W(q) = \tilde V_{\!\rm p-h}(q) + \tilde w_I(q) =  -\tF(q)\lambda(q)
  \tilde \Gamma_{\!\rm dd}(q)\,,
\end{equation}
along with the Bethe-Goldstone equation (\ref{eq:BGSchq})
  \begin{eqnarray}
    &&2\left[
      \tF(q)\lambda(q)\left[\sqrt{1+\Gamma_{\!\rm dd}(r)}-1\right]^{\cal F}(q)
      \right]^{\cal F}(r)
    \nonumber\\
    &&= -(v(r) + w_I(r)) \sqrt{1+\Gamma_{\!\rm dd}(r)}\,.
  \end{eqnarray}
Multiplying the latter equation with $\sqrt{1+\Gamma_{\!\rm dd}(r)}$ and 
combining it with the former yields the expression
  \begin{widetext}
  \begin{eqnarray}
    V_{\!\rm p-h}(r) &=& [1+\Gamma_{\!\rm dd}(r)]v(r) + \Gamma_{\!\rm dd}(r)w_{\rm I}(r)
    \\
    &-&\left[\tF(q)\lambda(q)\tilde\Gamma_{\!\rm dd}(q)\right]^{\cal F}(r)
   + 2\sqrt{1+\Gamma_{\!\rm dd}(r)}\left[
      \tF(q)\lambda(q)\left[\sqrt{1+\Gamma_{\!\rm dd}(r)}-1\right]^{\cal F}(q)
      \right]^{\cal F}(r)\,.\nonumber
  \end{eqnarray}
  \end{widetext}

Note that if we have $\tF(q)=t(q)$ and $\lambda(q)=1$, the terms on
the second line combine to
$\frac{\hbar^2}{m}\left|\nabla\sqrt{1+\Gamma_{\!\rm
    dd}(\rvec)}\right|^2$. Since $\SF(q) = 1$ for $q>2\KF$ and
$\lambda(q)\rightarrow 1$ for large $q$, and never differs from 1 by
more than 20 percent, the use of
$\frac{\hbar^2}{m}\left|\nabla\sqrt{1+\Gamma_{\!\rm
    dd}(\rvec)}\right|^2$ seems to be justified.
\section{State-dependent correlations}

\subsection{Operator structure}

In this paper we focus on interactions of the so-called $v_6$ form, which
in neutron matter involves only the first three operators
spelled out in Eq.~(\ref{eq:operator_v6}), \ie
\begin{equation}
  \hat v(r) =  v_c(r)\1 + v_\sigma(r){\bm\sigma}_1\cdot{\bm\sigma}_2 + v_S(r)
  S_{12}(\hat\rvec)\,.
  \label{eq:Vop1}
\end{equation}
An alternative choice of the interaction operators is
\cite{FNN82,SmithSpin}
\begin{equation}
 \hat v(r) =  v_c(r)\1 + v_L(r)\hat L(\hat\rvec) +  v_T(r)\hat T(\hat\rvec)
\end{equation}
where
\begin{equation}
  \hat L(\hat\rvec) \equiv ({\bm\sigma}_1\cdot \hat\rvec)({\bm\sigma}_2\cdot\hat\rvec)\,,
  \quad
  \hat T(\hat\rvec) \equiv {\bm\sigma}_1\cdot{\bm\sigma}_2-
  ({\bm\sigma}_1\cdot \hat\rvec)({\bm\sigma}_2\cdot\hat\rvec)
\end{equation}
are the ``longitudinal'' and ``transverse'' operators.
These operators are amenable to summations of RPA-type diagrams because 
they have the features
\begin{equation}
  \Tr_{{\bm\sigma}_3} \hat O_i(1,3)\hat O_j(3,2) = 2
  \hat O_i(1,2)\delta_{ij},\quad
  \hat O_i(1,2) \in \{\1,\hat L, \hat T\}\,.
\end{equation}

A third useful set of operators are the projectors
\begin{eqnarray}
  \hat P_{s\phantom{+}} &=& \frac{1}{4}\left(\1-{\bm\sigma}_1\cdot{\bm\sigma}_2\right)
  \nonumber\\
  \hat P_{t+} &=& \frac{1}{6}\left(3\,\1+\sij12+S_{12}(\hat\rvec)\right)
  \label{eq:projectors}\\
  \hat P_{t-} &=& \frac{1}{12}\left(3\,\1+\sij12-2S_{12}(\hat\rvec)
  \right)\,.
  \nonumber
  \end{eqnarray}
These satisfy the relations $\hat P_i(12) \hat P_j(12) = \hat P_i(12)
\delta_{ij}$ and $\hat P_1 + \hat P_2 + \hat P_3 = \1$ and are
therefore appropriate for solving the coordinate-space equations.
\begin{widetext}
The three sets of operators are related through
\begin{alignat}{3}
  \left(\begin{matrix}\1 \cr \sij{1}{2}\cr S_{12}(\hat\rvec)\end{matrix}
    \right)
  =& \qquad\left(\begin{matrix}1 &0 &\phantom{-}0 \cr 0 & 1 &
    \phantom{-}1\cr 0 & 2 & -1\end{matrix}\right)
  \left(\begin{matrix}\1 \cr \hat{L} \cr \hat{T}\end{matrix}\right)
  &&=
  \phantom{\frac{1}{12}}\left(\begin{matrix}\phantom{-}1 & 1 &
      \phantom{-}1 \cr -3 & 1 & \phantom{-}1\cr
     \phantom{-}0 & 2 & -4\cr\end{matrix}\right)
  \left(\begin{matrix}\hat{P}_{s\phantom{-}} \cr \hat{P}_{t+} \cr \hat{P}_{t-}\cr\end{matrix}
  \right)\,,\nonumber\\
  \left(\begin{matrix}\1 \cr \hat L \cr \hat T \end{matrix}\right)\hspace{0.3cm}
  =& \quad\frac{1}{3}\left(\begin{matrix}3 &0 &\phantom{-}0 \cr 0 &  1& \phantom{-}
    1
    \cr 0 & 2 & -1\cr\end{matrix}\right)
  \left(\begin{matrix}\1\cr \sij12 \cr S_{12}(\hat\rvec)\end{matrix}\right)
  &&=
  \phantom{\frac{1}{12}}
  \left(\begin{matrix}\phantom{-}1 & 1 & \phantom{-}1 \cr -1 & 1 & -1\cr
     -2 & 0 & \phantom{-}2\cr\end{matrix}\right)
     \left(\begin{matrix}\hat P_{s\phantom{-}} \cr \hat P_{t+} \cr \hat P_{t-}\cr\end{matrix}
     \right)\,,\\
     \left(\begin{matrix}\hat P_{s\phantom{-}} \cr \hat P_{t+} \cr \hat P_{t-}\cr\end{matrix}
     \right)\hspace{0.1cm}
     =& \quad\frac{1}{4}\left(\begin{matrix}
       1 &-1 &-1 \cr 2 &  \phantom{-}2& \phantom{-}
    0
    \cr 1 & -1 & \phantom{-}1\cr\end{matrix}\right)
     \left(\begin{matrix}\1 \cr \hat L \cr \hat T \end{matrix}\right)
  &&=
     \frac{1}{12}
     \left(\begin{matrix}
       3 & -3 & \phantom{-}0 \cr 6 & \phantom{-}2 & \phantom{-}2\cr
         3 & \phantom{-}1 & -2\cr\end{matrix}\right)
     \left(\begin{matrix}\1\cr \sij12 \cr S_{12}(\hat\rvec) \end{matrix}\right)
   \,.\nonumber
\end{alignat}
\end{widetext}

\subsection{Momentum space equation}

Particle-hole matrix elements are best calculated in the operator basis
$\{\1,{\bm\sigma}_1\cdot\bm{\sigma}_2,S_{12}(\hat\rvec)\}$, where for 
$O_1(1,2)$ to $O_4(1,2)$ we have
  \begin{eqnarray}
    &&\bra{\hvec+\qvec,\hvec'-\qvec}v_\alpha(1,2)O_\alpha(1,2)
    \ket{\hvec,\hvec'}\\
    &=& \frac{\rho}{N}\int d^3r v_\alpha(r)j_0(qr) O_\alpha(1,2)\,\qquad (1\le\alpha\le 4)\nonumber
  \end{eqnarray}
whereas we have for tensor components we have
  \begin{eqnarray}
    &&\bra{\hvec+\qvec,\hvec'-\qvec}_\alpha(1,2) O_\alpha(1,2,\hat\rvec_{12})
    \ket{\hvec,\hvec'}\\
    &=& -\frac{\rho}{N}\int d^3r v_\alpha(r)j_2(qr)
    O_\alpha(1,2,\hat\qvec)\,\qquad (5\le\alpha\le 6)\nonumber\,.
  \end{eqnarray}
Since there is no ambiguity, we will refer to both the $j_0$ Fourier
transform and the $-j_2$ Fourier transform by the tilde symbol defined
in Eq. (\ref{eq:ft}).
  
The momentum-space equation (\ref{eq:SRPA}) is best solved in the basis
$\{\1,\hat L,\hat T\}$, where we simply get the response function and 
the static structure function in the above operator channels, \ie
 \begin{eqnarray}
    \chi_\alpha(q,\omega) &=&
    \frac{\chi_0(q,\omega)} {1-\tilde V_{\rm
        p-h}^{(\alpha)}(q)\chi_0(q,\omega)}\label{eq:chiRPAalpha}\,,\\
 S_\alpha(q) &=& -\int_0^\infty \frac{d\omega}{\pi} \Im \chi_\alpha(q,\omega)
\label{eq:SRPAalpha}
\end{eqnarray}
with $\alpha = 1, 3, 5.$. These equations are identical to those of
Ref.~\onlinecite{SmithSpin} for bosons when the bosonic particle-hole
propagator (\ref{eq:Chi0Bose}) is inserted, whereas they give
Eq.~(\ref{eq:FermiPPA}) in the above three channels if the
``collective'' Lindhard function (\ref{eq:Chi0Coll}) is used.  For the
full Lindhard function, the integral must be carried out numerically.
Likewise, both the energy-dependent and the energy-independent
effective interactions $\widetilde W^{(\alpha)}(q,\omega)$ and $\widetilde
W^{(\alpha)}(q)$, as well as the induced interaction
(\ref{eq:Fermiwind}), are obtained in this way, using Eq.~(\ref{eq:Scond}).
Of course, the tensor operator introduces an angular dependence.

\subsection{Coordinate space equation}
Owing to the projection property $\hat P_i(12) \hat P_j(12) =
\hat P_i(12) \delta_{ij}$, the coordinate-space equations are best
formulated in the projector basis $\{\hat P_s, \hat P_{t+}, \hat
P_{t-}\}$. The only new aspect is that we must keep the angular
dependence of the tensor correlations in the kinetic energy term
$\left|\nabla\sqrt{1+\Gamma_{\!\rm dd}(\rvec)}\right|^2$.  For
its evaluation, let
\begin{equation}
  \sum_{i=1}^3(1+\Gamma_{\rm dd}^{(i)}(r))\hat P_i \equiv
  \left[\sum_{i=1}^3 f_i(r)\hat P_i\right]^2
  = \sum_{i=1}^3 f_i^2(r)\hat P_i\,.
\end{equation}
Then
\begin{eqnarray}
  \left|\nabla \sum_{i=1}^3 f_i(r)\hat P_i\right|^2 &=&
  \sum_{i=1}^3 \left|\frac {df_i(r)}{dr}\right|^2 \hat P_i
  - \frac{f_S^2(r)}{\hbar^2 r^2}\left|\Lvec S_{12}(\hat\rvec)\right|^2\,,
  \nonumber
\end{eqnarray}
where $f_S(r) = (f_{t+}(r)-f_{t-}(r))/6$ is the component of $\hat f(\rvec)$ in
the tensor channel. The last term is simplified using
\begin{eqnarray}
  \left|\Lvec S_{12}(\hat\rvec)\right|^2&=& \frac{1}{2}
  \left[\Lvec^2 S_{12}^2(\hat\rvec)-
    S_{12}(\hat\rvec)\left(\Lvec^2 S_{12}(\hat\rvec)\right)\right.\nonumber\\
    &&\left.-\left(\Lvec^2 S_{12}(\hat\rvec)\right)S_{12}(\hat\rvec)\right]\,.
\end{eqnarray}
Next, we invoke $\Lvec^2 S_{12}(\hat\rvec) = 6\hbar^2 S_{12}(\hat\rvec)$
and
\begin{equation}
  S_{12}^2(\hat\rvec) = (2\hat P_{t+}-4\hat P_{t-})^2 = 4\hat P_{t+}+16 P_{t-}\,,
  \end{equation}
leading to
\begin{eqnarray}
  \left|\Lvec S_{12}(\hat\rvec)\right|^2 &=&-6\hbar^2 S_{12}(\hat\rvec)-6\hbar^2
  S^2_{12}
  (\hat\rvec) \nonumber\\
  &=&-36\hbar^2 \hat P_{t+} - 72\hbar^2\hat P_{t-} \,.
\end{eqnarray}
We arrive at
\begin{eqnarray}
  \left|\nabla \sum_{i=1}^3 f_i(r)P_i\right|^2 &=&
  \sum_{i=1}^3 \left|\frac{d f_i(r)}{dr}\right|^2\hat P_i\nonumber\\
  &+& \frac{36}{r^2}f_S^2(r)\hat P_{t+} +  \frac{72}{r^2}f_S^2(r)\hat P_{t-}\,.
\label{eq:TensorKin}
\end{eqnarray}
Thereby, we have determined the structure of the particle-hole
interaction in the three projector channels. We can now go back to
the $\{\1,\hat L,\hat T\}$ basis and in momentum space and
solve the equations iteratively.

\subsection{Exchange corrections}
\label{ssec:exchanges}

\begin{figure}[H]
    \centerline{\includegraphics[width=0.6\columnwidth]{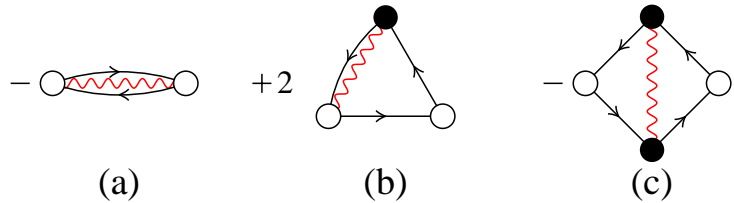}}
    \caption{The figure shows the diagrammatic representation of the
      lowest order exchange corrections $V_{\rm ee}(r)$ and $X_{\rm
        ee}(r)$. For the interaction correction $V_{\rm ee}(r)$, the
      red wavy line is to be interpreted as the effective interaction
      $W(r_{ij})$. In the correlation correction $X_{\rm ee}(r)$, the
      wavy red line represents the function $\Gamma_{\rm dd}(r)$.
       \label{fig:eelink}}
\end{figure}

Exchange diagrams have important consequences for the effective
interactions, particularly in nucleonic systems.  They must be
included even at low densities to achieve consistency between the
energetics and the quasiparticle interaction \cite{fullbcs}.  The
simplest version of the FHNC hierarchy that corrects for this
deficiency is FHNC//1, which includes the sum of the three exchange
diagrams shown in Fig. \ref{fig:eelink}.

The relevant modification from the full FHNC-EL equations as
formulated in Ref.~\citenum{polish} involves keeping only the exchange term
$V_{\rm ee}(k)$.  The Euler equation becomes 
\begin{equation}
     S(q) = \frac{\SF(q) + \tilde X_{\rm ee}(q)}
     {\sqrt{1+\displaystyle\frac{2\SF^2(q)}{t(q)}\tilde V_{\rm p-h}(q)}}\,.
       \label{eq:Stemp}
   \end{equation}
where the particle-hole interaction is modified by
\begin{equation}
  \tilde V_{\rm p-h}(q) \rightarrow \tilde V_{\rm p-h}(q) +
  \tilde V_{\rm ex}(q)\,,\quad \tilde V_{\rm ex}(q)\equiv
  \frac{\tilde V_{\rm ee}(q)}{\SF^2(q)}
  \label{eq:Vphexc}
\end{equation}
and where $ X_{\rm ee}(r_{12})$ and $V_{\rm ee}(r_{12})$ are given
by the sum of the three diagrams shown in Fig. \ref{fig:eelink}.

We have shown in Ref.~\onlinecite{fullbcs} that na\"\i ve addition of
exchange diagrams is problematic because it leads to an
incorrect low-density limit of the pair correlations.
We have rectified this situation by a slight modification of the
Euler equation, namely
\begin{equation}
 S(q) =
 \SF(q)\sqrt{\frac{1+\displaystyle\frac{2\SF^2(q)}{t(q)}\tilde V_{\rm
       ex}(q)} {1+\displaystyle\frac{2\SF^2(q)}{t(q)}\tilde
       V_{\rm p-h}(q)}}\,.
   \end{equation}
The square-root term in the numerator may be identified with a
``collective RPA'' expression for the exchange contribution to the
static structure function (for state-independent interactions this is
equal to the spin-structure function),
   \begin{equation}
     S_{\rm ex}(q) = \frac{\SF(q)}{\sqrt{1+\displaystyle\frac{2\SF^2(q)}{t(q)}
         \tilde V_{\rm ex}(q)}}\,,\label{eq:SsigmaColl}
   \end{equation}
The expression (\ref{eq:Stemp}) is then obtained by
expanding $S_{\rm ex}(q)$ to first order in the interactions and
identifying
   \[\tilde X_{\rm ee}(q) \approx -\frac{\SF^3(q)}{t(q)} \tilde V_{\rm ee}(q)\,.\]

We have commented above on the fact that, with the qualification that 
the Jastrow-Feenberg wave function is not exact, the positivity of 
the term under the square root in the denominator is related to the 
stability against density fluctuations. Likewise, the positivity of 
the numerator is connected with the stability against spin-density 
fluctuations.

In time-dependent Hartree-Fock theory \cite{ThoulessBook}, the
diagrams shown in Fig.~\ref{fig:eelink} correspond to the
particle-hole ladder diagrams, driven by the {\em exchange\/} term of
the particle-hole interaction
\begin{equation}
  W_{\rm ex}(\hvec,\hvec';\qvec)
  = \Omega\bra{\hvec+\qvec,\hvec'-\qvec} W \ket{\hvec',\hvec}\,.
  \label{eq:Wex}
\end{equation}
This non-local term supplements the RPA sum by the RPA-exchange (or
particle-hole ladder) summation.  The connection to the (local) FHNC
expression (\ref{eq:Vphexc}) is made by realizing that this expression
is obtained from the exact expression (\ref{eq:Wex}) by exactly the
same hole-state averaging process as was introduced in
Eq.~(\ref{eq:favg}):
\begin{equation}
  V_{\rm ex}(q) = \frac{\tilde V_{\rm ee}(q)}{S_{\rm F}^2(q)}
  = \left\langle  W_{\rm ex}(\hvec,\hvec';\qvec)\right\rangle(q)\,.
\end{equation}
For state-dependent correlations and interactions, we can simply go
back to the definition (\ref{eq:Wex}) and interpret the interaction
$W(r)$ as an operator of the form (\ref{eq:Vop1}). The calculation for
the central and spin components go exactly as before.  The tensor
component needs special treatment which will be outlined in the
appendix \ref{app:exchanges}.

\subsection{Energy\label{sec:fenergy}}

In calculating the energy, we can again simply follow the analysis of
Smith and Jackson, inserting exchange corrections where appropriate.
We must keep in mind that there is no finite truncation scheme of the
FHNC equations such that acceptable expressions for the pair
distribution function and the static structure function are the
Fourier transforms of each other.  That is, having obtained an
optimized static structure function $S(q)$, one must construct the
pair distribution function $g(r)$ by appropriate combination of
correlation diagrams and exchanges. In the case of state-independent
correlations, the simplest expression for $g(r)$ is
\begin{eqnarray}
  g(r) &=& \left[1+\Gamma_{\!\rm dd}(r)\right]\left[g_{\rm F}(r) + C(r)\right]\\
  \tilde C(q) &=& \left[\SF^2(q)-1\right]\tilde \Gamma_{\!\rm dd}(q)
  + (\Delta \tilde X_{\rm ee})(q)\,.
  \label{eq:gofr}
\end{eqnarray}
where $g_{\rm F}(r) = 1-\frac{1}{2}\ell^2(r\KF)$, with $\ell(x) =
3j_1(x)/x$ the pair distribution function of non-interacting
fermions. In the FHNC//1 approximation, $\SF(q)$ is replaced by
$\SF(q) + \tilde X_{\rm ee}(q)$, and $(\Delta \tilde X_{\rm ee})(q)$,
which is represented by the sum of diagram (b) and (c) shown in
Fig.~\ref{fig:eelink} is added to $\tilde C(q)$. Summarizing, we
obtain
\begin{eqnarray}
  \frac{E}{N}
  &=&\frac{T_{\rm F}}{N} + \frac{E_{\rm R}}{N} +
  \frac{E_{\rm Q}}{N}
\,,\nonumber \\
\frac{E_{\rm R}}{N} &=& \frac{\rho }{2}\int\! d^3r\>
\bigl[g_{\rm F}(r) + C(r)\bigr]\biggl[(1+\Gamma_{\!\rm dd}(r))v(r)\nonumber\\
 && \qquad + \frac{\hbar^2}{m}\left|\nabla\sqrt{1+\Gamma_{\!\rm dd}(r)}\right|^2\biggr]\,,
\label{eq:ER}\\
\frac{E_{\rm Q}}{N} &=& \frac{1}{4}\int\!\frac{d^3q}{(2\pi)^3\rho}\>
t(q)\tilde\Gamma_{\!\rm dd}^2(q)\left[S^2_{\rm F}(q)/S(q)-1\,\right]\,,
\nonumber\\
\label{eq:EQ}
\end{eqnarray}
where $T_{\rm F}$ is the kinetic energy of the non-interacting Fermi gas.
In the state-dependent case, $g(r)$ becomes an operator
in spin-space,
\begin{eqnarray}
  \hat g(r) &=& \left[1+\Gamma_{\!\rm dd}^{(s)}(r)\right]
    \left[1+C_s(r)-\ell^2(r\KF)\right]\hat P_s\nonumber\\
    &+& \left[1+\Gamma_{\!\rm dd}^{(t+)}(r)\right]
    \left[1+C_{t+}(r)+\ell^2(r\KF)\right]\hat P_{t+}\nonumber\\
    &+&\left[1+\Gamma_{\!\rm dd}^{(t-)}(r)\right]
    \left[1+C_{t-}(r)+\ell^2(r\KF)\right]\hat P_{t-}\nonumber\\
    &\equiv& g_s(r) \hat P_s + g_{t+}(r)\hat P_{t+}(r) + g_{t-}(r)\hat P_{t-}(r)
\,
\end{eqnarray}
with which we obtain the potential energy
\begin{eqnarray}
  &&\frac{\left\langle \hat V \right\rangle}{N}
  = \frac{\rho}{2}\Tr\int d^3r \hat v(r)\hat g(r)\nonumber\\
  &=&\frac{\rho}{4}\int d^3r \left[v_s(r) g_s(r) + 2v_{t+}(r)g_{t+}(r)\right.\nonumber\\
   & &\left.+ v_{t-}(r)g_{t-}(r)\right]\nonumber
\end{eqnarray}
The kinetic energy term in Eq.~(\ref{eq:ER}) is generalized to 
state-dependent correlations in the same way, without the 
$[1+\Gamma_{\!\rm dd}^{(\alpha)}(r)]$ factors.  Note, of course, 
that we need to keep the kinetic-energy correction spelled out in
Eq.~\ref{eq:TensorKin}. Finally, the term $E_{\rm Q}$ is generalized to
\begin{eqnarray}
  \frac{E_{\rm Q}}{N} &=& \frac{1}{4}\int \frac{d^3q}{(2\pi)^3\rho} t(q)
  \sum_\alpha (\tilde\Gamma_{\!\rm dd}^{(\alpha)})^2(q)\left[S^2_{\rm F}(q)/S_\alpha(q)-1\,\right]\times\nonumber\\
  &&\times\Tr O_{\alpha}^2(1,2)\,\qquad O_\alpha \in \{\hat 1,\hat L, \hat T\}\,.
\end{eqnarray}

\section{Results} 

In applying the EL-FHNC procedures established in preceding sections,
we have chosen as inputs the $v_6$ truncation of the Reid interaction
as formulated in Ref.~\onlinecite{Day81} and the Argonne $v_{6}'$
interaction \cite{AV18}.  For each of these interactions, we have
performed a sequence of computations.  In terms of correlation
operators $\Gamma_{\!\rm dd}^{(\alpha)}(r)$, (i) keeping only central
components, (ii) including both central and spin operators, and (iii)
supplementing the latter with tensor operators, in each case omitting
or keeping the exchange diagrams described in Section
\ref{ssec:exchanges}.  Additionally, we have used the ``collective
approximation'' (\ref{eq:Chi0Coll}) as well as the exact Lindhard
function in both the calculation of $S(q)$ by means of
Eq.~(\ref{eq:SRPA}) and the calculation of the effective interaction
through Eq.~(\ref{eq:Scond}).  As is usual in FHNC notation, we
designate the level at which exchange diagrams are included by //n,
e.g., //0 means no exchanges are included, while //1 means that we
keep the one-line diagrams $X_{\rm ee}^{(1)}(r)$ and $V_{\rm
  ee}^{(1)}(r)$. Calculations using the ``collective approximation''
will be referred to as ``FHNC'' and those containing the exact
particle-hole propagator, as ``parquet''.

\subsection{Energetics}

We shall refrain here from showing the large array of results obtained
in the FHNC and parquet versions of the theory and focus on the most 
telling implementations.  Our calculations have been extended
to much lower densities than is usually done \cite{Baldo2012}, since 
the regime of very low density has been of recent interest due to
the expectation there is some fundamental similarity between low-density 
neutron matter and the unitary gas. Among other significant features, 
the superfluid gap at low densities is close to 0.5 
times the Fermi energy \cite{Gezerlis2014,GC2008,ectpaper}.  However,
we did not go quite as low in density as in our previous work, since 
good resolution in both coordinate and momentum space would require a
much larger mesh.

\begin{figure}[H]
\centerline{\includegraphics[width=0.7\columnwidth,angle=-90]{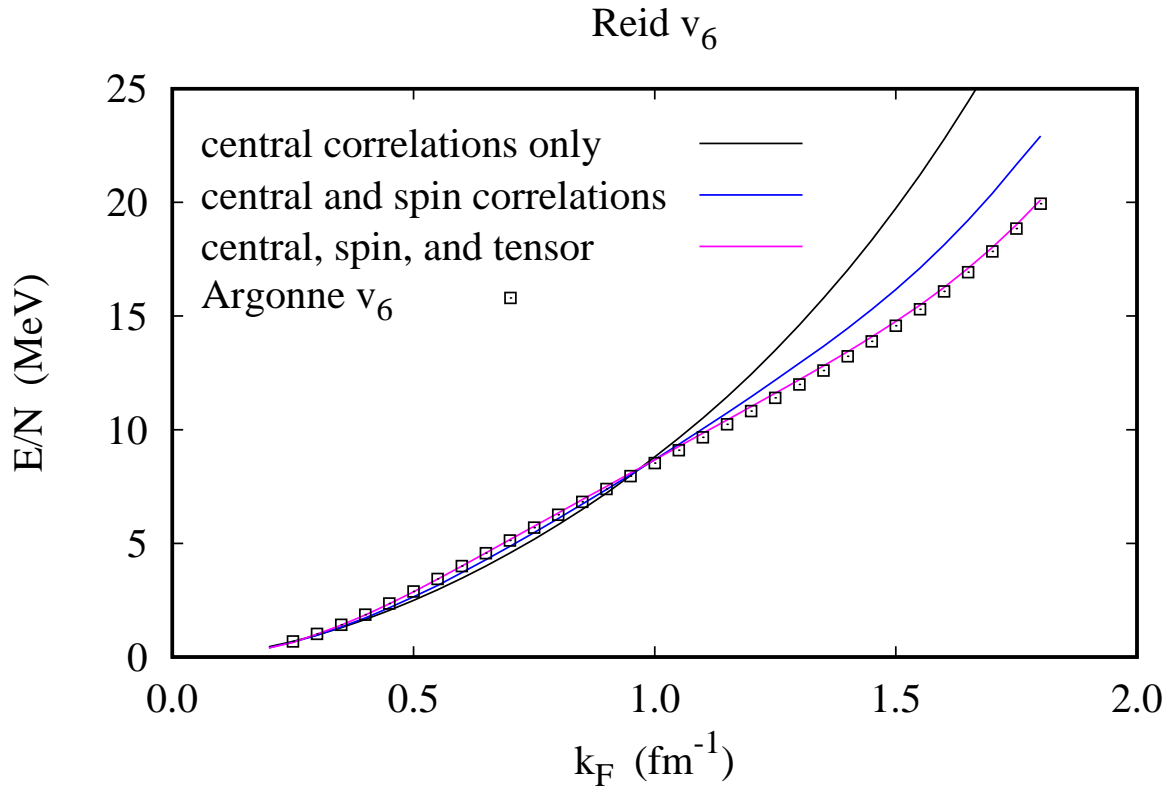}}
\caption{(color online) The neutron-matter equation of state for the 
Reid $v_6$ potential is plotted versus Fermi wave number $\KF$ in three 
approximations: (i) accounting only for central correlations (black line), 
(ii) including both spin and central correlations (blue line), and (iii) 
further introducing tensor correlations (magenta line).  Also shown, for 
the third case of central, spin, and tensor correlations, are 
results for the Argonne $v_6$ potential (boxes).
\label{fig:re_eosplot}}
\end{figure}

The first quantity of interest is, of course, the energy per particle, 
with results exhibited in Fig.~\ref{fig:re_eosplot}.
Shown there are only the calculations containing exchange diagrams and
the full particle-hole propagator, the plots for other calculations 
being omitted for clarity.

\begin{figure}[H]
\centerline{\includegraphics[width=0.7\columnwidth,angle=-90]{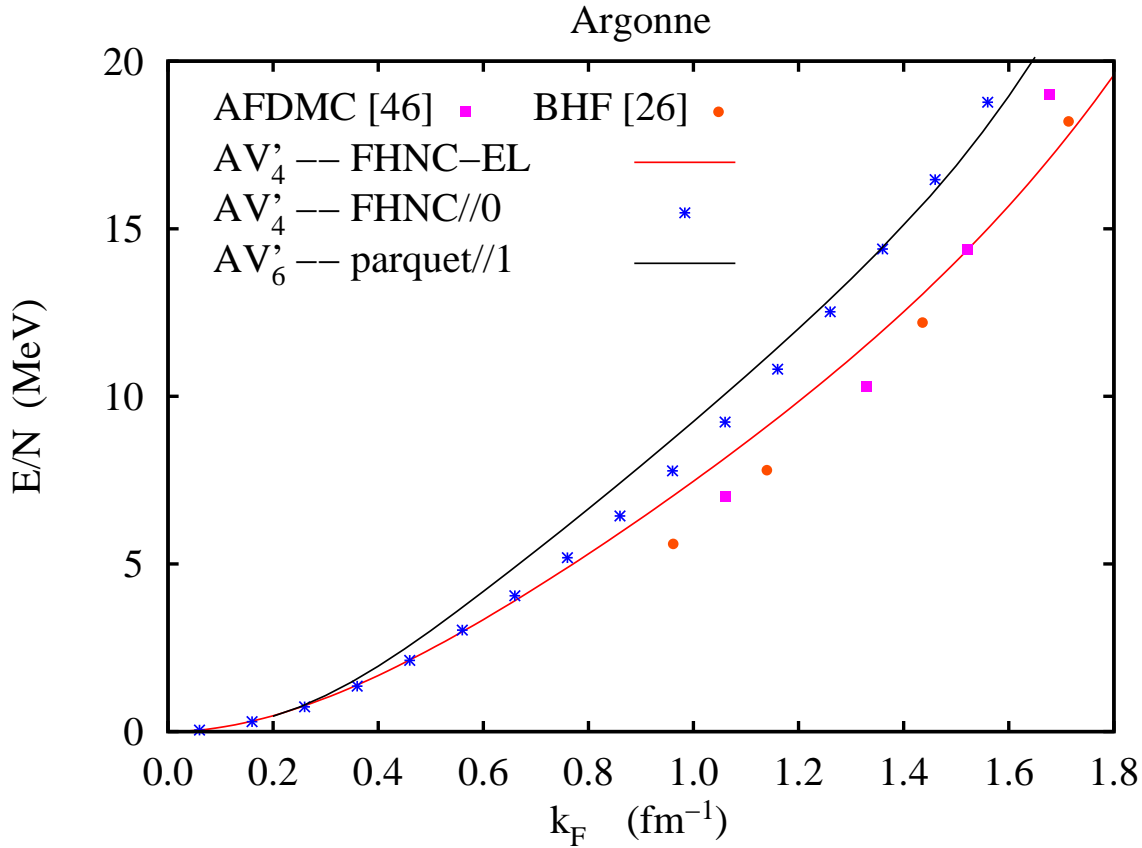}}
\caption{(color online) The figure shows a comparison of neutron
  matter equation of state for the Argonne $v_4'$ and $v_6'$
  interactions for the state-independent FHNC-EL and FHNC//0 calculations of
  Ref. \onlinecite{ectpaper}, the present work, as well as
  from the auxiliary-field diffusion Monte
  Carlo (AFDMC) method \cite{Gandolfi2009a} for the Argonne $v_{18}$
  interaction and from a Brueckner-Hartree-Fock
  calculation \cite{Baldo2012} for the Argonne $v_4'$ potential.
        \label{fig:ar_eosplot}}
\end{figure}

We observe that the equations of state begin to differ visibly beyond
$\KF =\, 1{\rm fm}^{-1}$, we note, of course, that at that density the
FHNC//0 and FHNC//1 approximations deviate from a full FHNC-EL
calculation by about the same amount, see Fig. \ref{fig:ar_eosplot}
and Fig. 1 of Ref.~\onlinecite{ectpaper}.  In fact, in view of the
difference in the correlation functions found in different
approximations to be discussed below we find the agreement between
different calculations rather remarkable. We also direct the attention
to the fact that the results for the Argonne potential are rather
close to those of the Reid interaction.

As stated above, we keep the comparison with earlier calculations to a
minimum because extensive work is avaliable \cite{Baldo2012}. Fig.
\ref{fig:ar_eosplot} gives an update for two versions of the Argonne
potential \cite{AV18} including the calculations of this work, the
state-independent full FHNC-EL calculations of
Ref. \onlinecite{ectpaper}, the Brueckner-Hartree-Fock calculations of
Baldo \etal, and the auxiliary-field diffusion Monte Carlo (AFDMC)
method \cite{Gandolfi2009a}.

\begin{figure}[H]
\centerline{\includegraphics[width=0.7\columnwidth,angle=-90]%
{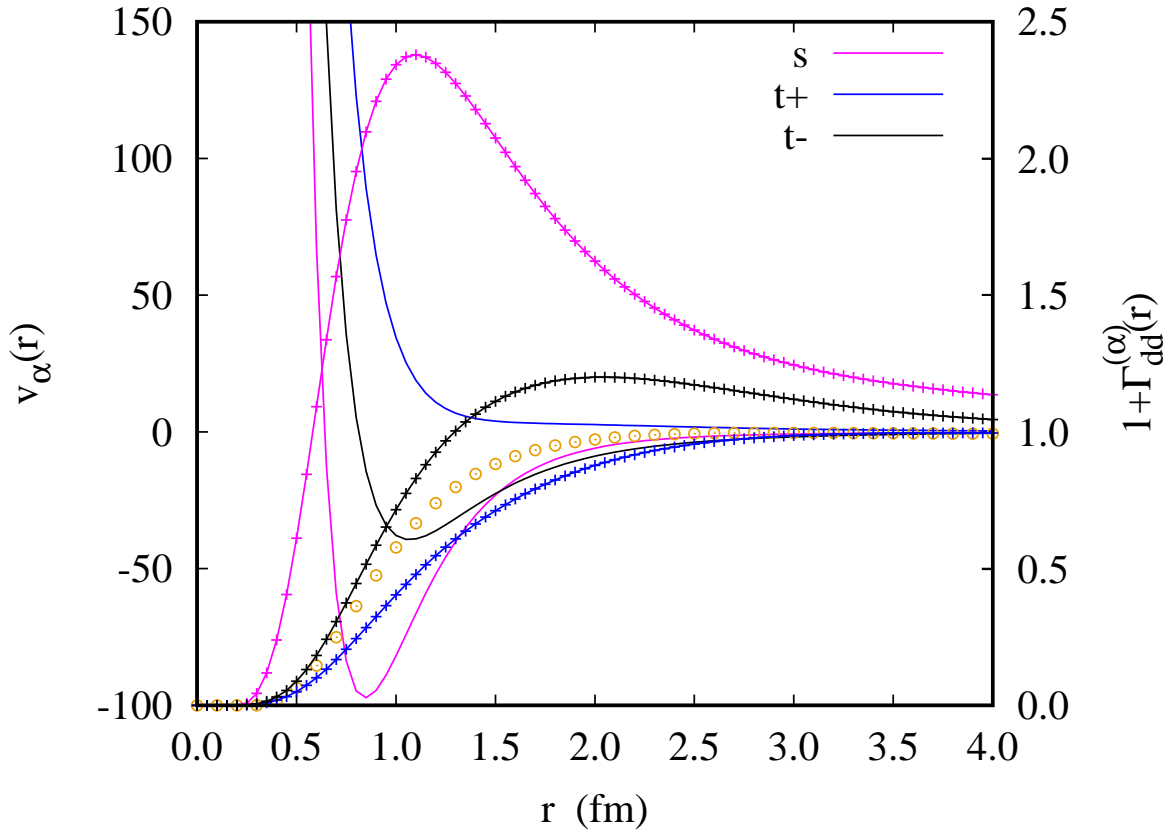}}
\caption{(color online) Plots, versus radial separation $r$, of the 
components of the Reid $v_6$ interaction in spin-triplet states and 
of the corresponding projectors $\hat P_s$, $\hat P_{t+}$, and $\hat P_{t-}$ 
(magenta, blue, and black), together with the dynamic correlation functions
$1+\Gamma_{\!\rm dd}^{(\alpha)}(r)$ in these channels (same colors, 
lines with ``$+$'' markers), all at Fermi wave number $\KF =\, 1\,{\rm
fm}^{-1}$.  Also shown is the correlation function $1+\Gamma_{\!\rm dd}(r)$ 
for state-independent correlations (yellow circles).
\label{fig:re_gddplot}}
\end{figure}

The close similarity of the energetics exhibited in Fig.~\ref{fig:re_eosplot} 
for the three quite different calculations is, however, by no means 
an indication that central correlations are sufficient for a description
of this system, as was demonstrated in another sequence of calculations.  
Fig.~\ref{fig:re_gddplot} shows the bare interactions in the three projector 
channels $\hat{P}_s$, $\hat{P}_{t+}$, and $\hat{P}_{t-}$, along with the 
dynamic correlation functions $1+\Gamma_{\!\rm dd}^{(\alpha)}(r)$ in these 
channels. Obviously the interactions are very different; for 
example, recall that the $S$-wave interaction has a scattering length of 
$a_0 \approx -18.7\ $fm \cite{PhysRevLett.83.3788}, \ie it is close to 
developing a bound state. Correspondingly, the correlation develops a 
strong peak roughly at the location of the interaction minimum.  The 
two triplet channels are much less attractive; in fact the $t+$ channels 
is repulsive, hence the particles tend to be pushed apart.

We conclude this section with a brief comparison with other many-body
approaches; a very extensive comparison of numerical data from
different approaches may be found in Ref.~\onlinecite{Baldo2012}.

Variational and perturbative calculations are often referred to as
complimentary approaches.  We feel that this view is somewhat
oversimplified: It was already observed by Sim {\em et al.}
\cite{Woo70}, and re-iterated by Jackson {\em et al.}
\cite{parquet1,parquet2,parquet3} that the boson Bethe-Goldstone
equation is indeed a proper subset of the calculation within the
optimized HNC scheme. We have clarified above and in
Ref.~\onlinecite{fullbcs} to what extent the same is true for
fermions.  We have come to the conclusion that the only additional
postulate is that the pair wave function is a function of the
interparticle distance; see section \ref{ssec:ladders}. In other
words, for fermions the Brueckner-Hartree-Fock (BHF) theory is also a
proper subset of FHNC-EL. The essential difference is that in
conventional (BHF), {\em ad-hoc\/} constraints must be introduced to
prevent the pair wave function from becoming unphysically long ranged
\cite{BetheBrandowPetschek}.  Exactly the same is true when a
variational theory is truncated at low order: The Euler-equation in
2-body approximation has unphysically long-ranged solutions that must
be somehow tamed; this is done, for example, by the so-called
``low-order constrained variational (LOCV) method.''  In the FHNC-EL
scheme, the ``induced interaction'' $w_I(r)$ makes sure that the
long-ranged behavior of the correlations is physically reasonable;
artificially imposed constraints are therefore unnecessary.

\subsection{Correlation and Distribution functions}

Two questions are addressed in this subsection: The first is what it takes
to have a reliable prediction for the distribution and structure functions,
and the second, once that is determined, how physical quantities of
interest depend on density and specific features of the interaction. 

\begin{figure}[H]
\centerline{\includegraphics[width=0.7\columnwidth,angle=-90]%
{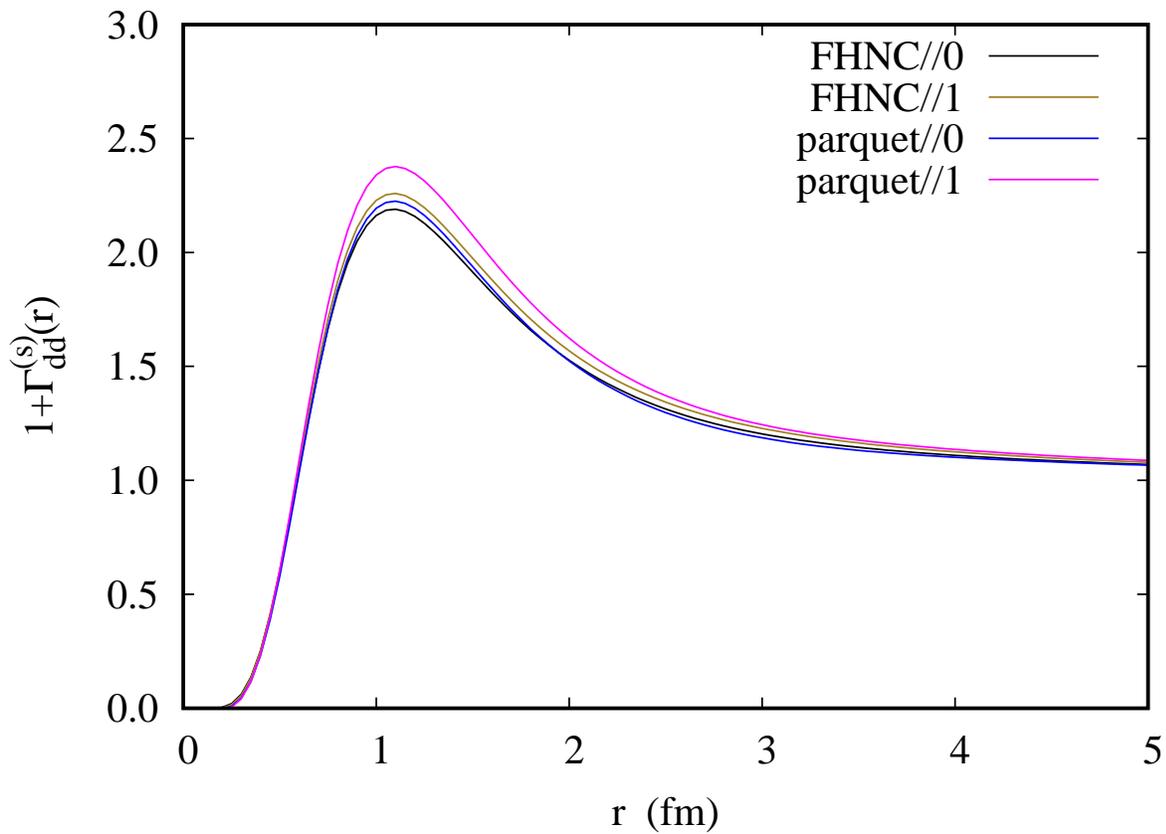}}
\caption{(color online) Plots, at Fermi wave number $\KF =\, 1\, 
{\rm fm}^{-1}$, of the spin-singlet dynamic correlation function
$1+\Gamma_{\!\rm dd}^{(s)}(r)$ versus radius $r$, as obtained in 
the four different approximations explained in the text.
\label{fig:re_gdd_s}}
\end{figure}

\begin{figure}[H]
  \centerline{\includegraphics[width=0.7\columnwidth,angle=-90]%
    {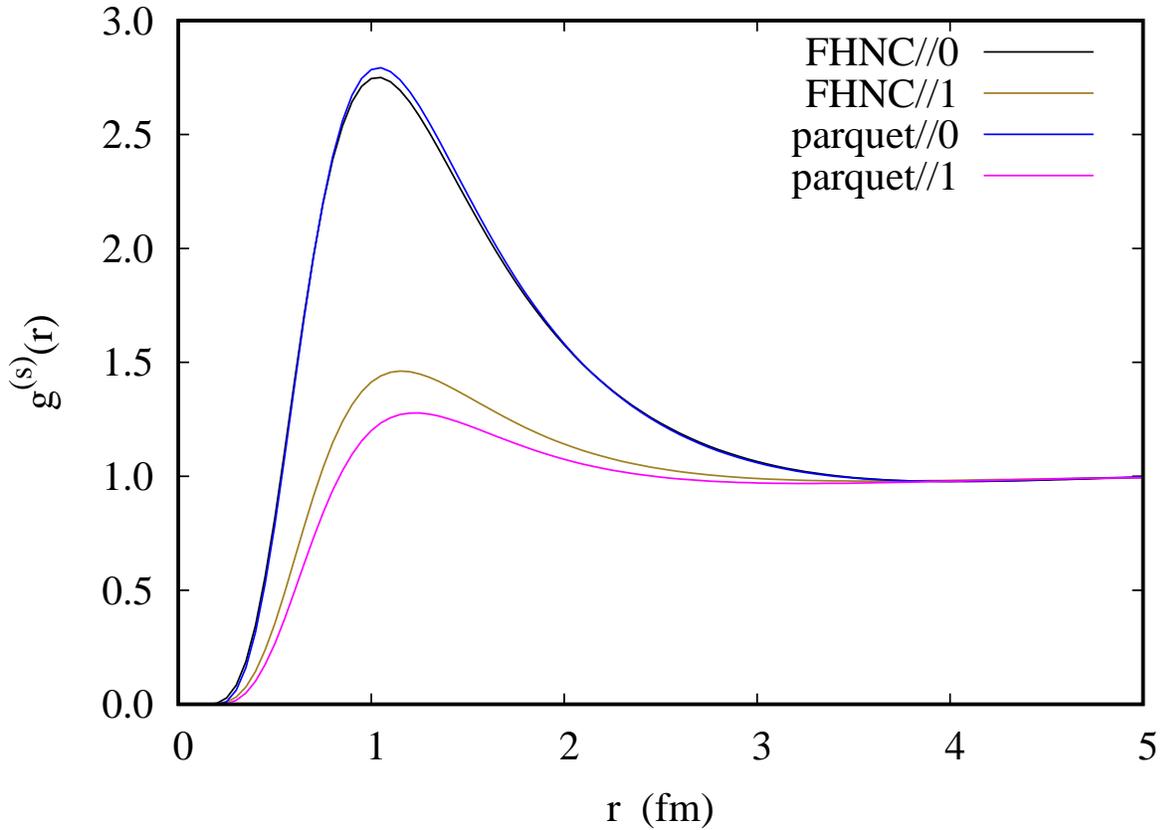}}
    \caption{(color online) This figure shows, at $\KF =\, 1\, {\rm
    fm}^{-1}$, the spin-singlet pair distribution function in 
     the four different approximations as explained in the text.
     \label{fig:re_gofr_s}}
\end{figure}

We have partly addressed the first issue already in the preceding
subsection, where we have shown that simple state-independent correlations
can reproduce the energetics with reasonable accuracy, but they give
no reliable prediction for the dynamic correlations. The other question
is concerned with the importance of exchange diagrams and propagator
corrections. We address this question partly in
Fig.~\ref{fig:re_gdd_s} and Fig.~\ref{fig:re_gofr_s}, where we show the
dynamic correlation function and the pair distribution function in the
singlet channel in four different approximations: without and with the
exchange contribution and propagator corrections (labeled FHNC//0 and
FHNC//1), as well as with propagator corrections (labeled parquet//0
and parquet//1). Evidently, all of these corrections have little
consequence for the direct dynamic correlations.

The situation changes remarkably for the pair distribution function
(Fig.~\ref{fig:re_gofr_s}), where we see that exchanges have a rather
drastic effect. We hasten to explain that this is exclusively due to
the term $(\Delta \tilde X_{\rm ee})(q)$ spelled out in
Eq.~(\ref{eq:gofr}); the replacement $\SF(q) \rightarrow\SF(q) +
(\Delta \tilde X_{\rm ee})(q)$ has a negligible effect. It was observed a long time ago that the
sum of the three diagrams shown in Fig.~\ref{fig:eelink} is much smaller
than the three individual terms \cite{Kro77};
the fact that the individual terms are quite large is peculiar
to the present situation.  These terms are relatively small in
\he3 and low-density gases.

Let us finally turn to the density dependence of the correlations.  We
have already pointed out that the $s$-channel is close to forming a
bound state; accordingly, we find that the $s$-projection of the correlation
function develops a strong nearest-neighbor peak as the density
decreases. With increasing density, this peak is suppressed, evidently
by both the Pauli principle and the induced interaction $w_{\rm I}(r)$.

\begin{figure}[H]
  \centering
\subfigure{
  \includegraphics[width=0.42\columnwidth,angle=-90]{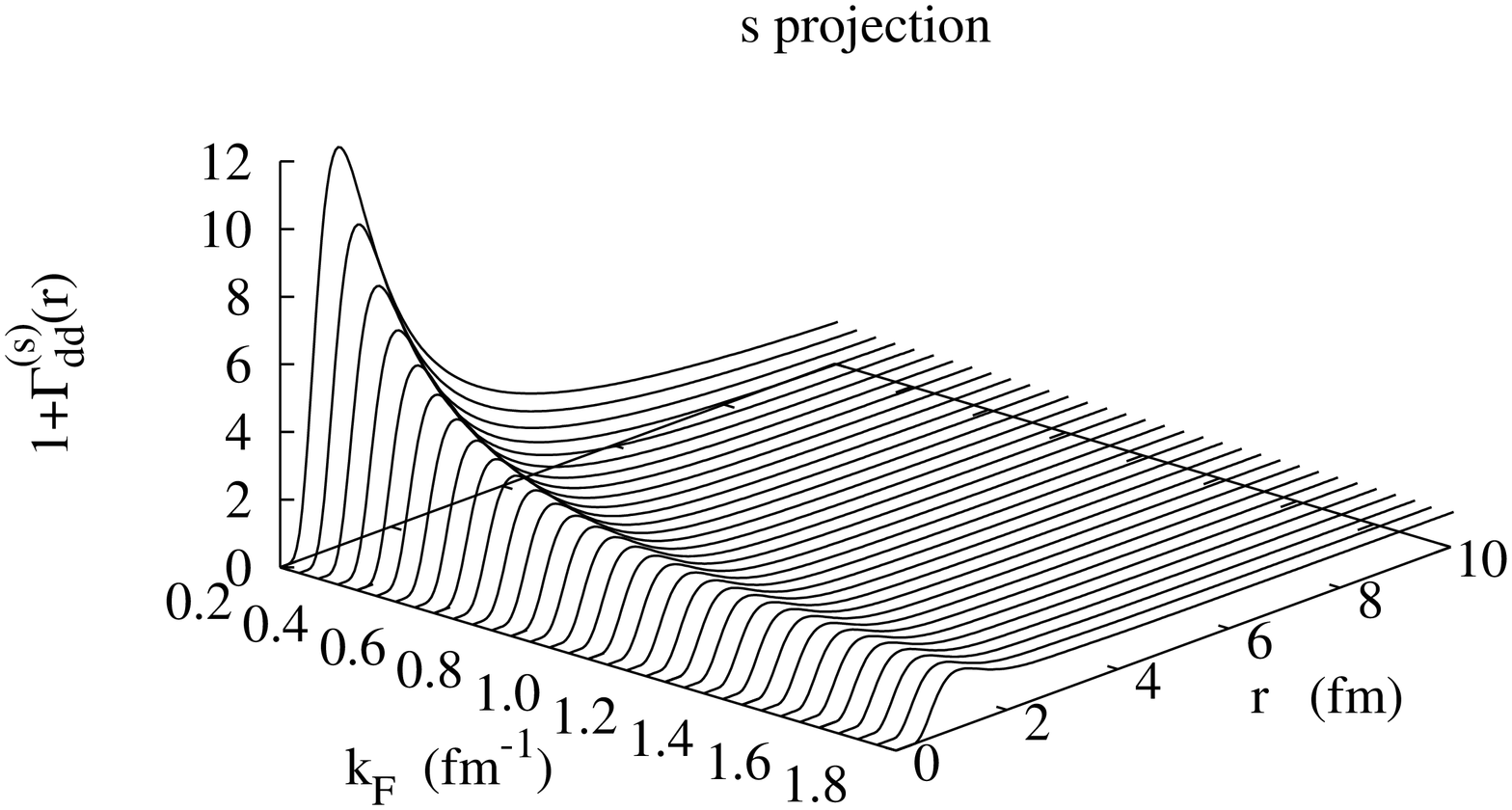}
  \label{fig:gdds}
}
\subfigure{
  \includegraphics[width=0.42\columnwidth,angle=-90]{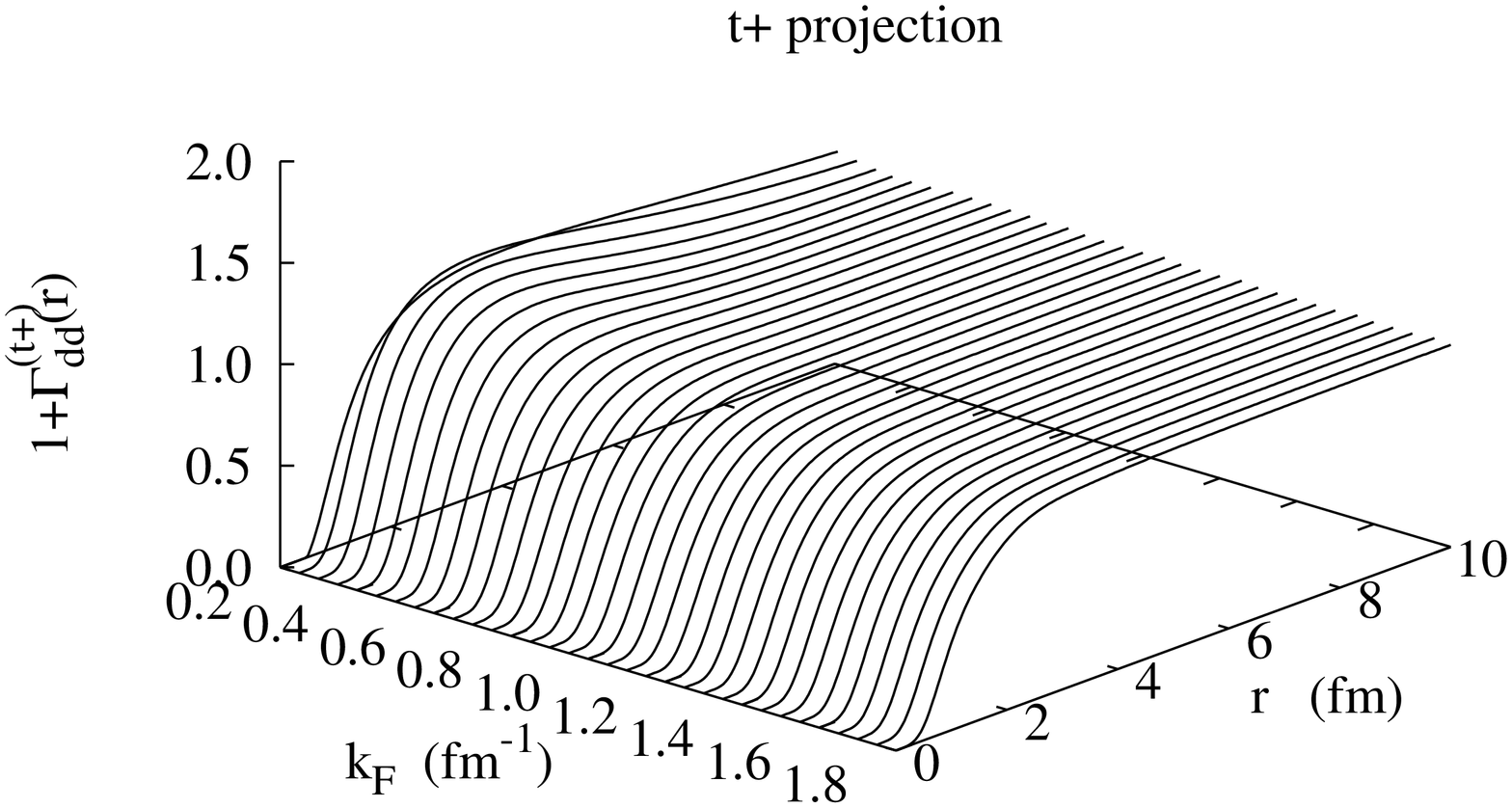}
  \label{fig:gddtp}
}
\subfigure{
  \includegraphics[width=0.42\columnwidth,angle=-90]{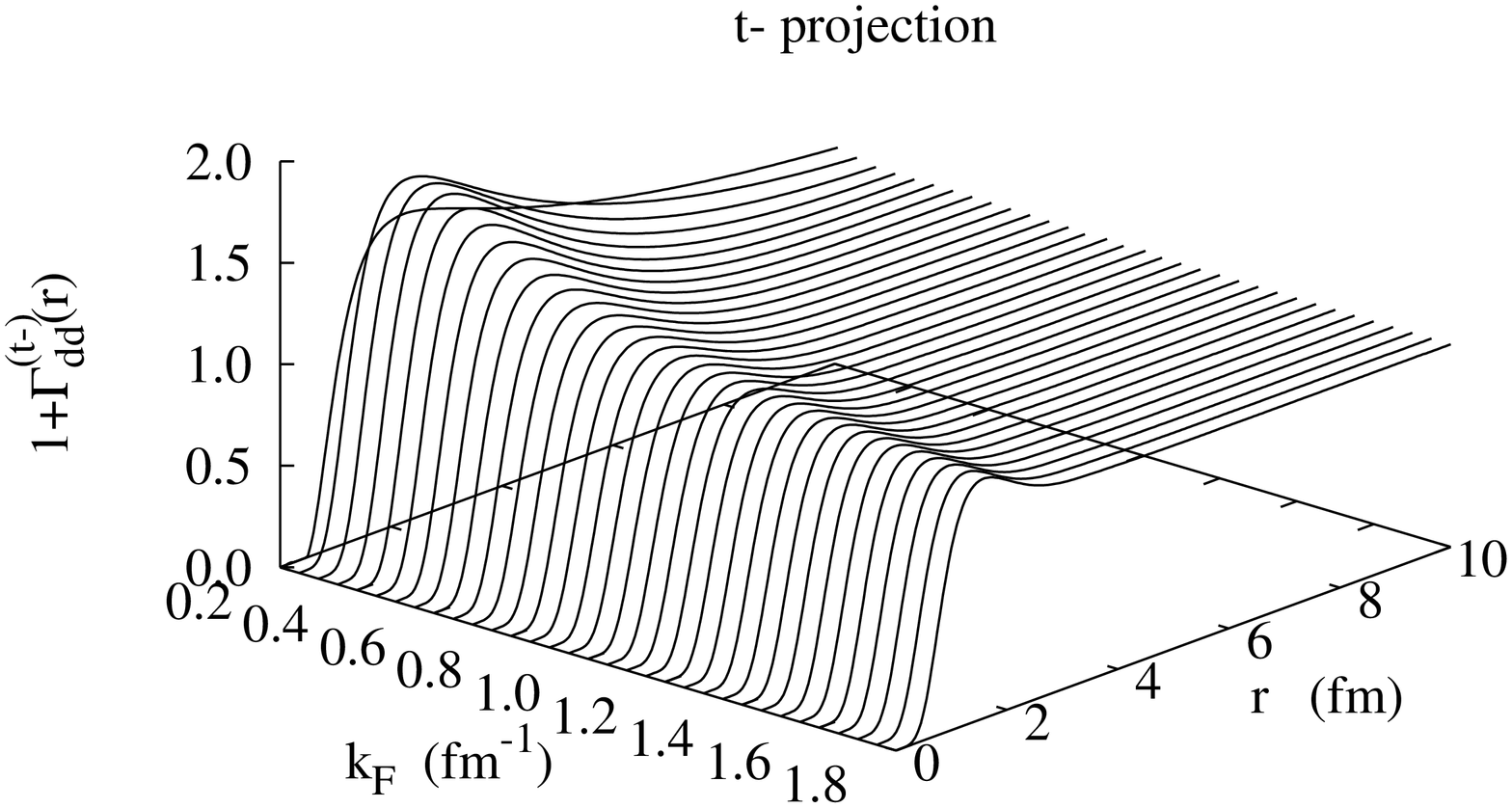}
  \label{fig:gddtm}
  }
\caption{The figure shows the density dependence of the
  dynamic correlation functions $1+\Gamma_{\!\rm dd}^{(\alpha)}(r)$
  in the three spin-projector channels $s$ (Fig.~\ref{fig:gdds}),
  $t+$ (Fig.~\ref{fig:gddtp}), and  $t-$ (Fig.~\ref{fig:gddtm}), 
  \label{fig:re_Gdd_3d}}
\end{figure}
  
\subsection{Effective interactions}

Effective interactions are perhaps more relevant than ground state
properties because they determine quantities like the response
\cite{Wam93} and pairing properties \cite{SedrakianClarkBCSReview}
that are directly observable. We address here again the same two
questions that we have posed above: What is an acceptable
computational procedure to determine these interactions, and how do
they depend on external parameters like the density ?

The most important input for linear response theory and, hence for the
calculation of the dynamic structure function, is the particle-hole
interaction. The long-wavelength limit of the particle-hole interaction
is related to the hydrodynamic speed of sound by
\begin{equation}
mc^2 = \frac{d}{d\rho}\rho^2 \frac{d}{d\rho}\frac{E}{N}\,.
\label{eq:mcfromeos}
\end{equation}
In a Fermi fluid, we also have Pauli repulsion, reflected in the 
relation
\begin{equation}
mc^2 = mc_{\rm F}^{*2} + \tilde V_{\rm p-h}(0+) \equiv  mc_{\rm F}^{*2}(1+F_0^S)
\,,
\label{eq:FermimcfromVph}
\end{equation}
where $c_{\rm F}^* = \sqrt{\frac{\hbar^2\KF^2}{3mm^*}}$ is the speed of
sound of the non-interacting Fermi gas with the effective mass $m^*$,
and $F_0^s$ is Landau's Fermi liquid parameter.  Requiring a positive
compressibility leads to Landau's stability condition $F_0^s > -1$.

The relationships (\ref{eq:mcfromeos}) and (\ref{eq:FermimcfromVph})
normally give identical predictions only in an exact theory
\cite{EKVar,parquet5}; good agreement is typically reached only at
very low densities. The reason for that is the very simple fact that
the convergence of cluster expansions for the Fermi-Liquid parameters
is intrinsically worse than that for the energy \cite{EKVar}: The
contribution to any $n$-body diagram to the energy is multiplied by
roughly a factor $n^2$ in an equivalent expansion of the
incompressibility from Eq.~(\ref{eq:mcfromeos}). Even in the much
simpler system $^4$He, where four- and five-body elementary diagrams
and three-body correlations are routinely included, the two
expressions (\ref{eq:mcfromeos}) and (\ref{eq:FermimcfromVph}) can
differ by up to a factor of 2 \cite{lowdens}.

The situation is even more complicated in Fermi systems due to the
multitude of exchange diagrams, of which we kept only the simplest.
Hence, one can expect good agreement only at very low densities
\cite{fullbcs}, but not at the densities considered here.

\begin{figure}[H]
  \centerline{\includegraphics[width=0.7\columnwidth,angle=-90]%
    {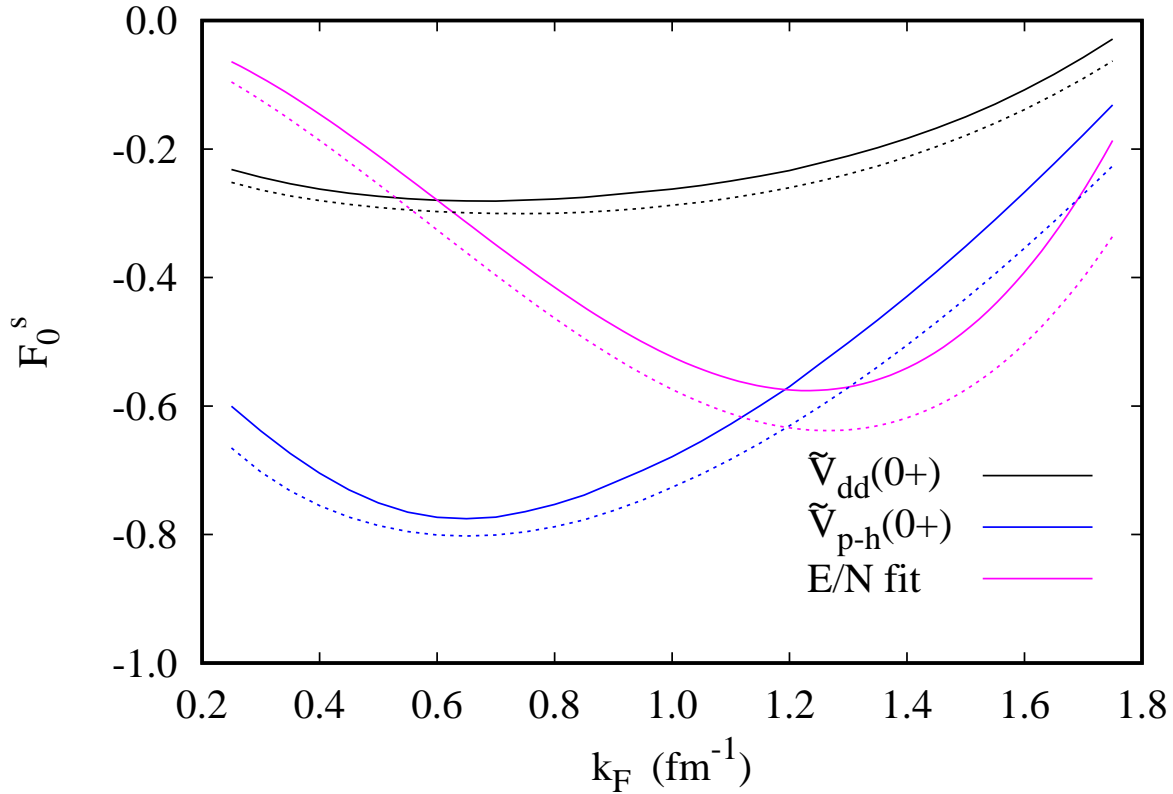}}
  \caption{(color online) This figure shows the density dependence of the
    Fermi liquid parameter $F_0^s$, as obtained from the speed of
    sound via the derivative (\ref{eq:mcfromeos}) (magenta line) and
    from the long-wavelength limit (\ref{eq:FermimcfromVph}) of the
    microscopic particle-hole interaction. The black curve contains
    only the ``direct'' part (\ref{eq:Vph}), while the blue line
    contains both direct and exchange parts. The solid lines show 
    parquet//1 results, whereas the dashed lines show results from
    FHNC//1. \label{fig:re_f0s}}
\end{figure}

Fig.~\ref{fig:re_f0s} compares the results from
Eqs.~(\ref{eq:mcfromeos}) and (\ref{eq:FermimcfromVph}) for the Reid
$v_6$ interaction and the parquet//1 calculation including tensor
correlations, shown as the magenta curve in Fig. \ref{fig:re_eosplot}.
The derivative (\ref{eq:mcfromeos}) was calculated by finite
differences, to eliminate numerical noise $m c^2$ has been fitted by a
third-order polynomial.

To connect microscopic and hydrodynamic speeds of sound we have
used an effective mass ratio $m^*/m=1$, which seems to be justified
by our results from Ref.~\onlinecite{fullbcs}. A number of
conclusions can be drawn from Fig.~\ref{fig:re_f0s}. The ``direct''
part is somewhat improved compared with the calculation based on
state-independent interactions, where the $F_0^s$ came out positive
(\cf Fig.~9 of Ref.~\onlinecite{fullbcs}). One would have expected
that $F_0^s$ goes to zero linearly as $\KF\rightarrow 0$; this appears
to happen only at much lower densities. The underlying cause seems to
be the strong density dependence of the singlet correlation functions
shown in Fig.~\ref{fig:gdds}. The state-independent calculation of
Ref.~\onlinecite{fullbcs} does not have this feature --  linear
behavior can be observed up to $\KF\ \overset{<}{\approx}\ 0.4\,{\rm
fm}^{-1}$. We also find that the contribution from exchange
diagrams is quite substantial. We attribute the remaining difference
partly to higher-order exchange diagrams, but also to the omission 
of ``elementary'' diagrams.

\begin{figure}[H]
  \centerline{\includegraphics[width=0.65\columnwidth,angle=-90]%
    {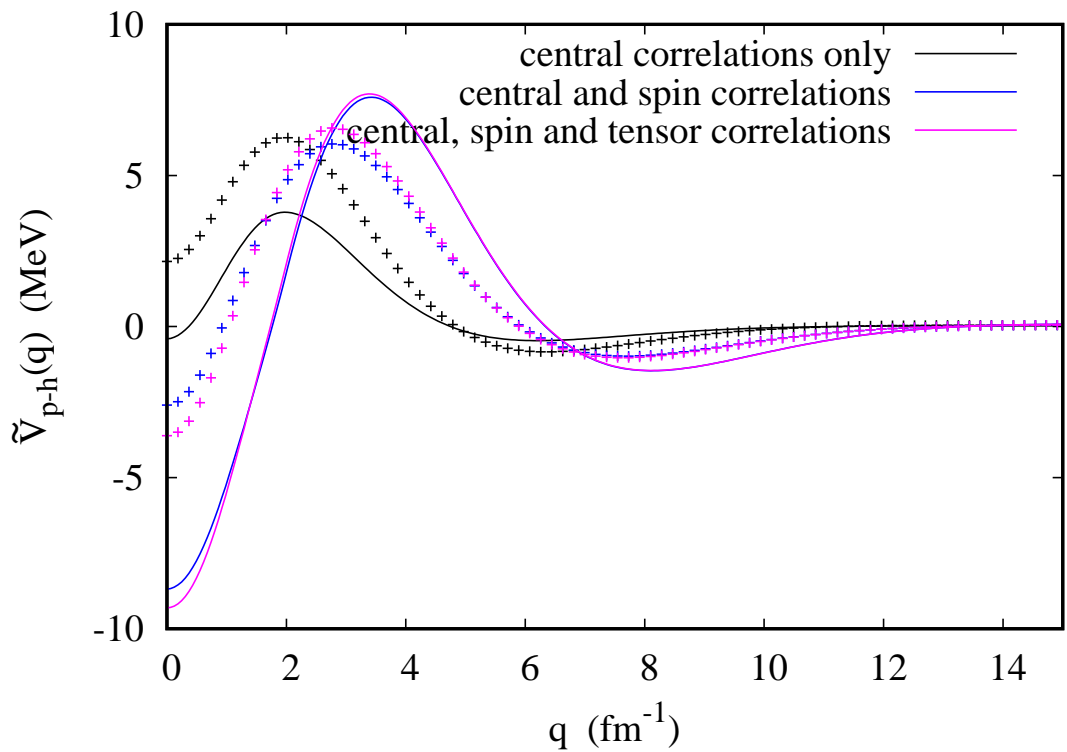}}
    \caption{(color online)  For $\KF =\, 1\, {\rm fm}^{-1}$, this
    figure depicts the central channel of the particle-hole interaction
    $\tilde V_{\!\rm p-h}(q)$ (Eq.~(\ref{eq:Vphexc})), based on
    (i) state-independent correlation functions (solid black line),
    (ii) spin-dependent correlations (blue line), and (iii) spin 
    and tensor correlations (magenta line).  Also shown is the direct part,
    given by Eq.~(\ref{eq:Vph}), in the three approximations 
    ($+$ symbols, same color). \label{fig:re_vph_c}}.
\end{figure}

In Fig.~\ref{fig:re_vph_c} we turn to the reliability of 
the derived particle-hole interaction
in successive approximations, \ie,  using central correlations,
both central and spin-dependent correlations, and finally adding 
tensor correlations. Evidently, central correlations cannot give 
a valid prediction of the particle-hole interaction.  Important 
corrections come from the contribution from exchange diagrams, 
$\tilde V_{\rm ex}(k)$.  This is to be expected and is consistent 
with our findings from Fig.~\ref{fig:re_f0s} and Ref.~\onlinecite{fullbcs}.  
Tensor correlations introduce some attraction at long wavelengths, 
but have minimal impact on the effective interaction in the 
central channel.

\begin{figure}[H]
\centering
\subfigure{
\includegraphics[width=0.42\columnwidth,angle=-90]{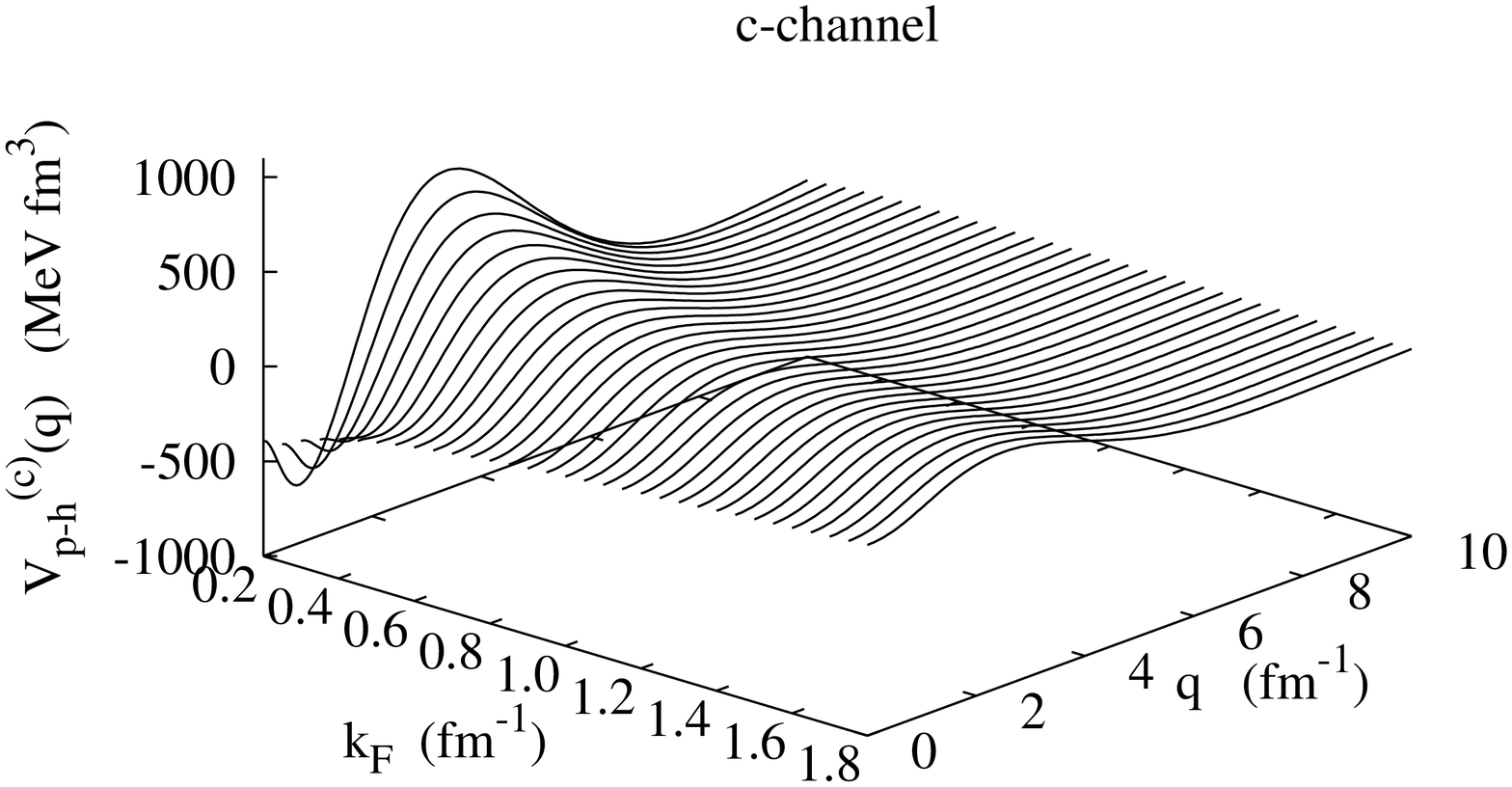}
\label{fig:Vphs}
}
\subfigure{
\includegraphics[width=0.42\columnwidth,angle=-90]{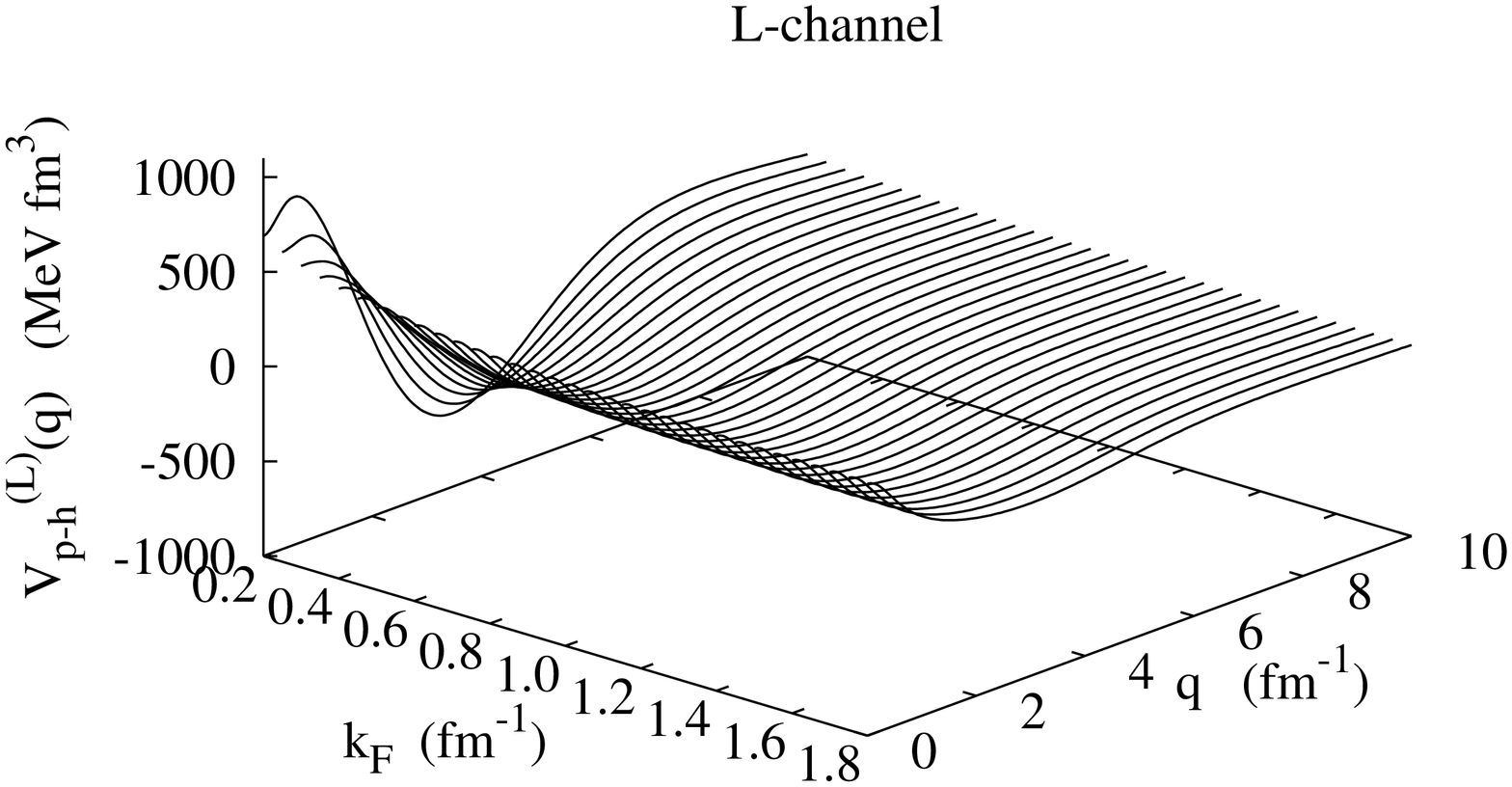}
\label{fig:VphL}
}
\subfigure{
\includegraphics[width=0.42\columnwidth,angle=-90]{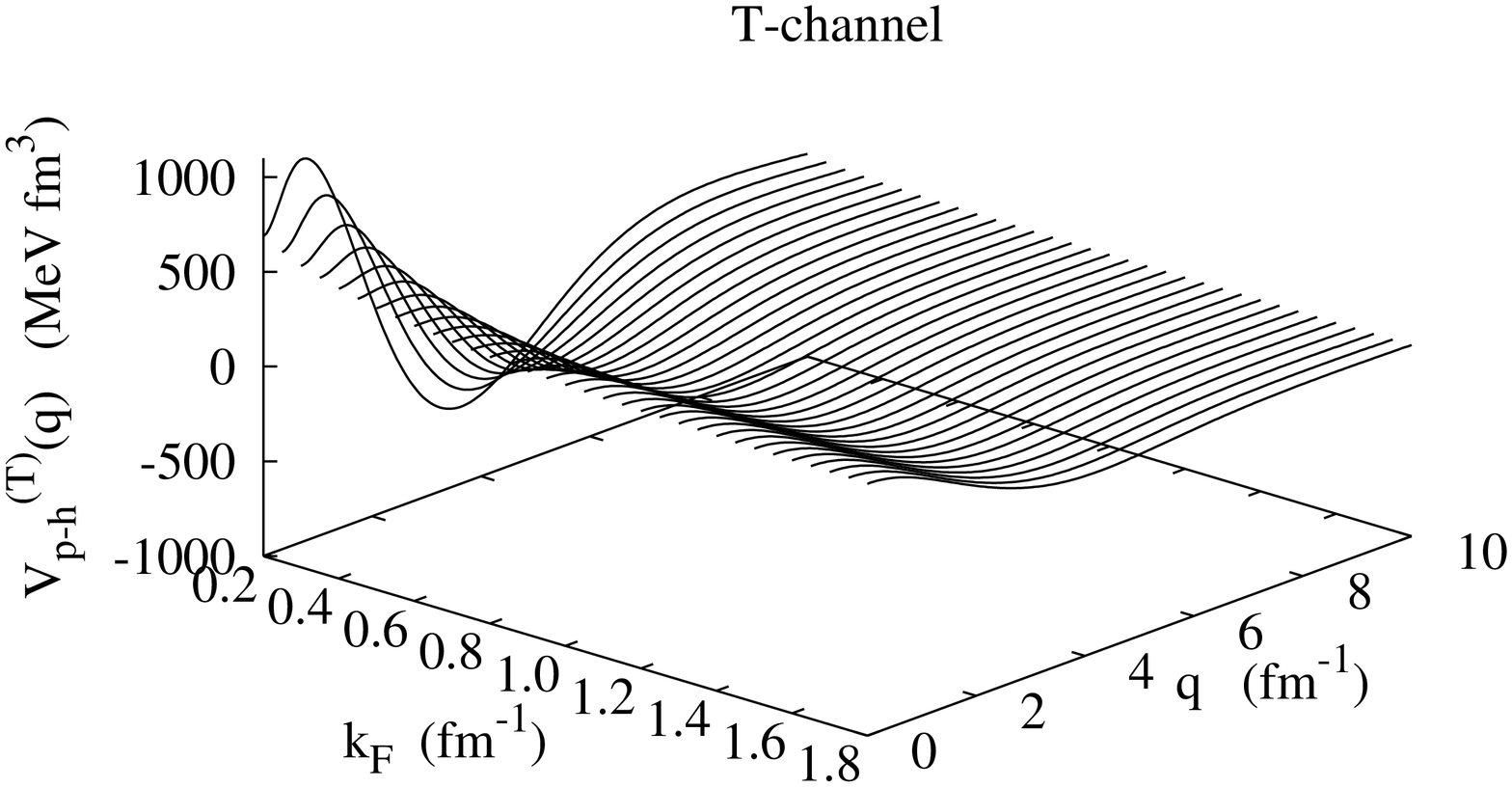}
\label{fig:VphT}
  }
\caption{This figure demonstrates the density dependence of the particle-hole
interaction $V_{\!\rm p-h}^{(\alpha)}(q)$ in the three operator
channels $\alpha=c$ (Fig.~\ref{fig:Vphs}), $L$ (Fig.~\ref{fig:VphL}), 
and $T$ (Fig.~\ref{fig:VphT}). Note that we show here 
$V_{\!\rm p-h}^{(\alpha)}(q) = \tilde V_{\!\rm p-h}^{(\alpha)}(q)/\rho$ 
in order to make the low-density behavior visible.
\label{fig:re_Vph_3d}}
\end{figure}

An overview of the density dependence of the effective interactions 
in the three channels $\{\1, \hat L, \hat T\}$ is provided in 
Figs.~\ref{fig:re_Vph_3d}. We emphasize, as discussed above, that 
only the central channel is attractive, whereas both the longitudinal 
and transverse channel interactions are repulsive.

Generally, we have not found a significant change of our results caused
by propagator corrections.  Approximating the response function by a
``collective'' version can be expected to be a good approximation 
only if the system has a strong collective mode. This is the case 
for both the transverse and the longitudinal channels, where
the interactions are repulsive. The most likely case where a
correction could be found would be the singlet channel because there
we have a negative $F_0^s$, which means that the zero-sound mode is
Landau damped.  We found, neverthelss, that the collective approximation
(\ref{eq:Chi0Coll}) in the RPA expression (\ref{eq:SRPA}) is very good, 
being most important when the system approaches the stability limit 
$F_0^s \rightarrow -1$.

The particle-hole interactions are necessary input for the calculation
of excitations, whereas the other effective interactions, $\hat
W_\alpha(r)$, are necessary for the calculation of pairing properties
\cite{HNCBCS,bcs,fullbcs}.  The latter differ from the particle-hole
interactions only by the induced interaction $w_{\rm I}(r)$; see
Eq.~(\ref{eq:BoseWind}).  In the present cases, we found that these
corrections are rather small and it would offer little further insight
to show them.

\section{Summary}

In this paper, we have taken up work by Smith and Jackson
\cite{SmithSpin} and generalized it to Fermi systems.  In doing so, we
have relied heavily on what is known from the optimized variational
(F)HNC method and made use of the fact that the parquet equations of
Ref.~\onlinecite{SmithSpin} could also be obtained from the bosonic
version of the (F)HNC equations of Fantoni and Rosati
\cite{FantoniSpins}.  In that way, and also by comparison with the
state-independent Jastrow-Feenberg case \cite{fullbcs}, we have been
able to identify the localization procedures leading from the general
parquet equations to the ``local'' ones. In the boson case, there was
just one such procedure, namely the approximation of the generally
energy-dependent induced interaction by an energy-independent form
(\ref{eq:Wlocal}), which holds for both bosons and fermions. This was
already recognized in Ref.~\onlinecite{mixmass}. In the case of
fermions, there is an additional localization procedure, which leads
to correlation functions depending only on the interparticle distance
(see Eq.~(\ref{eq:favg})), which turns the pair wave function of the
Bethe-Goldstone equation into a function of the interparticle
distance.

Our procedure goes beyond the so-called ``FHNC-SOC'' (single-operator-chain) 
approximation in the following sense.  The SOC approximation calculates 
only the operator-dependent chain diagrams, the correlation functions 
in these channels commonly being obtained by the LOCV procedure. In that 
sense, the SOC approximation may be understood as an RPA with an effective 
interaction \cite{rings}. We also sum these chain diagrams, but our 
correlation functions are determined, in all operator channels, by 
the localized Bethe-Goldstone equation. Moreover, the ``induced 
interactions'' also have the full operator structure.

The present implementation of the parquet theory does not solve the
notorious problem of the commutator diagrams. Model studies for a
fictitious system of bosons with spins \cite{SpinTwist} have indicated
that they can be very important if the interactions in spin-singlet
and spin-triplet cases are very different, which is indeed the case 
here, as indicated in Figs.~\ref{fig:re_Gdd_3d}. However, 
comparison with parquet theory should offer a much more elegant solution of the
problem of commutators than carrying out the symmetrization operators
for a variational wave function of the form (\ref{eq:f_prodwave}). The
next step in implementing the parquet strategy is to generalize the
Bethe-Goldstone equation to include all time-orderings of the induced
interaction $w_{\rm I}(r)$ in the summation of ladder diagrams, thereby 
superseding the Bethe-Goldstone equation. Work in this direction is
in progress.
\newpage

\appendix*
\section{Calculation of exchange diagrams}

\label{app:exchanges}

The easiest way to calculate the exchange corrections is to begin with
Eq.~(\ref{eq:Wex}):
\begin{equation}
 V_{\rm ex}(q) =
 \left\langle W_{\rm ex}(\hvec,\hvec',\qvec)\right\rangle\,.
\end{equation}
The input is normally a coordinate-space representation of $W(1,2)$;
in other words the effective interaction has the form
\begin{equation}
 \hat W(1,2) = W_0(r)\hat P_s + W_+(r)\hat P_{t+} + W_-(r)\hat P_{t-}\,.
\label{eq:WopP}
\end{equation}
The calculation of the central 
has been outlined in Ref.~\onlinecite{Kro77}; spin-correlations are dealt 
with in exactly the same way. The matrix elements of the tensor operator 
must be calculated independently.  A working formula for these exchange
diagrams is
\begin{eqnarray}
\tilde V_{\rm ee}(k) &=& - \frac{\rho}{\nu}\int d^3 r {\cal W}(\rvec)
\biggl[\ell^2(r\KF) j_0(rk) -\\
&& \ell(r\KF)(I(\kvec;\rvec)+I^*(\kvec;\rvec))+
I(\kvec;\rvec)I^*(\kvec;\rvec)\biggr]\,.\nonumber
\end{eqnarray}
Here $I(\kvec;\rvec)$ is conveniently calculated by an expansion in 
spherical harmonics, 
\begin{eqnarray}
I(\kvec;\rvec) &=& \frac{3}{4\pi\KF^3}\int d^3k'
e^{\I\kvec'\cdot\rvec}n(k')n(|\kvec-\kvec'|)\nonumber\\
&=&\sum_\ell(2\ell+1)
\I^\ell P_\ell(\cos(\hat\kvec\cdot\hat\rvec)) c_\ell(k,r)\,
\end{eqnarray}
with
\begin{equation}
 c_\ell(k,r) = \frac{3}{2{\KF}^3}\int_0^{\KF} dp p^2  j_\ell(rp)
\int_{x_L}^1 dx P_\ell(x)\,,
\end{equation}
where
\begin{equation}
x_L = \begin{cases} 1 &\mbox{if } |p-k| > \KF\\
 -1 &\mbox{if } p+k < \KF\\
\frac{p^2 + k^2 - \KF^2}{2pk}&\mbox{otherwise}\,.
\end{cases}
\end{equation}
{\em For central forces\/} this procedure gives
\begin{eqnarray} 
\tilde V_{\rm ee}(k)&=& -\rho\int d^3 r {\cal W}(r)
\biggl[\ell^2(r\KF) j_0(rk) - 2\ell(r\KF)c_0(k;r)\nonumber\\
&&\qquad + \sum_{\ell=0}^\infty(2\ell+1)c_\ell^2(k;r)\biggr]\,.
\end{eqnarray}
We have verified in Ref.~\onlinecite{Kro77} that
keeping $c_0(k,r)$ and $c_1(k,r)$ is generally sufficient.

For the tensor force, we obtain
\begin{eqnarray}
 \tilde V_{S,ij}(\kvec) &=& -\rho
 \int d^3 r {\cal W}_S(r)(3\hat x_i\hat x_j-\delta_{ij})
 \biggl[\ell^2(r\KF) e^{\I\kvec\cdot\rvec} \nonumber\\ 
 &&-\ell(r\KF)(I(\kvec;\rvec)+I^*(\kvec;\rvec))+
 I(\kvec;\rvec)I^*(\kvec;\rvec)\biggr]\,.\nonumber
\end{eqnarray}
The first term is just the $j_2$ Fourier transform.  The other two
terms can be calculated by expansion in spherical harmonics.  The
contributions from the $c_0(k,r)$ terms are zero due to the angle
integration. The only term that survives is the $c_1$ contribution
from the last term, which reads 
\begin{eqnarray} &&-9\rho\int d^3 r {\cal W}_S(r)
c_1^2(k,r)\sum_{ij}(3\hat x_i\hat x_j - \delta_{ij}) z^2\sigma_i\sigma_j
\nonumber\\
&=& -\frac{18}{15}\rho\int d^3 r {\cal W}_S(r)
          c_1^2(k,r)\left(2\sigma_z\sigma_z-\sigma_z\sigma_x-
          \sigma_y\sigma_y\right)\nonumber\\
          &=& -\frac{6}{5}\rho\int d^3 r {\cal W}_S(r)c_1^2(k,r)
          S_{12}(\hat\kvec)\,.
          \end{eqnarray}
If we keep only $c_0(k,r)$ and $c_1(k,r)$ we arrive at
\begin{equation}
          \tilde V_{\rm ex}[W_S] =
          \rho\int d^3 r {\cal W}_S(r)\left[
            \ell^2(r\KF) j_2(kr)
          - \frac{6}{5}
           c_1^2(k,r)\right]\,.
        \end{equation}
 

\begin{acknowledgments}
  This work was supported, in part, by the the College of Arts and
  Sciences of the University at Buffalo, SUNY. Encouragement for this
  work was derived from a workshop on {\em Nuclear Many-Body Theories:
    Beyond the mean field approaches\/} at the Asia Pacific Center for
  Theoretical Physics in Pohang, South Korea, in July 2019. We thank
  J. W. Clark for numerous comments and suggestions on this
  manuscript. One of us (JW) thanks the Austrian Marshall Plan
  Foundation for support during the summer 2018 and Robert Zillich for
  discussions.
\end{acknowledgments}

\newpage


\begin{thebibliography}{58}%
\makeatletter
\providecommand \@ifxundefined [1]{%
 \@ifx{#1\undefined}
}%
\providecommand \@ifnum [1]{%
 \ifnum #1\expandafter \@firstoftwo
 \else \expandafter \@secondoftwo
 \fi
}%
\providecommand \@ifx [1]{%
 \ifx #1\expandafter \@firstoftwo
 \else \expandafter \@secondoftwo
 \fi
}%
\providecommand \natexlab [1]{#1}%
\providecommand \enquote  [1]{``#1''}%
\providecommand \bibnamefont  [1]{#1}%
\providecommand \bibfnamefont [1]{#1}%
\providecommand \citenamefont [1]{#1}%
\providecommand \href@noop [0]{\@secondoftwo}%
\providecommand \href [0]{\begingroup \@sanitize@url \@href}%
\providecommand \@href[1]{\@@startlink{#1}\@@href}%
\providecommand \@@href[1]{\endgroup#1\@@endlink}%
\providecommand \@sanitize@url [0]{\catcode `\\12\catcode `\$12\catcode
  `\&12\catcode `\#12\catcode `\^12\catcode `\_12\catcode `\%12\relax}%
\providecommand \@@startlink[1]{}%
\providecommand \@@endlink[0]{}%
\providecommand \url  [0]{\begingroup\@sanitize@url \@url }%
\providecommand \@url [1]{\endgroup\@href {#1}{\urlprefix }}%
\providecommand \urlprefix  [0]{URL }%
\providecommand \Eprint [0]{\href }%
\providecommand \doibase [0]{http://dx.doi.org/}%
\providecommand \selectlanguage [0]{\@gobble}%
\providecommand \bibinfo  [0]{\@secondoftwo}%
\providecommand \bibfield  [0]{\@secondoftwo}%
\providecommand \translation [1]{[#1]}%
\providecommand \BibitemOpen [0]{}%
\providecommand \bibitemStop [0]{}%
\providecommand \bibitemNoStop [0]{.\EOS\space}%
\providecommand \EOS [0]{\spacefactor3000\relax}%
\providecommand \BibitemShut  [1]{\csname bibitem#1\endcsname}%
\let\auto@bib@innerbib\@empty
\bibitem [{\citenamefont {{Reid, Jr.}}(1968)}]{Reid68}%
  \BibitemOpen
  \bibfield  {author} {\bibinfo {author} {\bibfnamefont {R.~V.}\ \bibnamefont
  {{Reid, Jr.}}},\ }\href@noop {} {\bibfield  {journal} {\bibinfo  {journal}
  {Ann. Phys. (NY)}\ }\textbf {\bibinfo {volume} {50}},\ \bibinfo {pages} {411}
  (\bibinfo {year} {1968})}\BibitemShut {NoStop}%
\bibitem [{\citenamefont {Bethe}\ and\ \citenamefont
  {Johnson}(1974)}]{Bethe74}%
  \BibitemOpen
  \bibfield  {author} {\bibinfo {author} {\bibfnamefont {H.~A.}\ \bibnamefont
  {Bethe}}\ and\ \bibinfo {author} {\bibfnamefont {M.~B.}\ \bibnamefont
  {Johnson}},\ }\href@noop {} {\bibfield  {journal} {\bibinfo  {journal} {Nucl.
  Phys. A}\ }\textbf {\bibinfo {volume} {230}},\ \bibinfo {pages} {1} (\bibinfo
  {year} {1974})}\BibitemShut {NoStop}%
\bibitem [{\citenamefont {Day}(1981)}]{Day81}%
  \BibitemOpen
  \bibfield  {author} {\bibinfo {author} {\bibfnamefont {B.~D.}\ \bibnamefont
  {Day}},\ }\href@noop {} {\bibfield  {journal} {\bibinfo  {journal} {Phys.
  Rev. C}\ }\textbf {\bibinfo {volume} {24}},\ \bibinfo {pages} {1203}
  (\bibinfo {year} {1981})}\BibitemShut {NoStop}%
\bibitem [{\citenamefont {Wiringa}\ \emph {et~al.}(1995)\citenamefont
  {Wiringa}, \citenamefont {Stoks},\ and\ \citenamefont {Schiavilla}}]{AV18}%
  \BibitemOpen
  \bibfield  {author} {\bibinfo {author} {\bibfnamefont {R.~B.}\ \bibnamefont
  {Wiringa}}, \bibinfo {author} {\bibfnamefont {V.~G.~J.}\ \bibnamefont
  {Stoks}}, \ and\ \bibinfo {author} {\bibfnamefont {R.}~\bibnamefont
  {Schiavilla}},\ }\href@noop {} {\bibfield  {journal} {\bibinfo  {journal}
  {Phys. Rev. C}\ }\textbf {\bibinfo {volume} {51}},\ \bibinfo {pages} {38}
  (\bibinfo {year} {1995})}\BibitemShut {NoStop}%
\bibitem [{\citenamefont {Wiringa}\ \emph {et~al.}(1984)\citenamefont
  {Wiringa}, \citenamefont {Smith},\ and\ \citenamefont {Ainsworth}}]{Wiri84}%
  \BibitemOpen
  \bibfield  {author} {\bibinfo {author} {\bibfnamefont {R.~B.}\ \bibnamefont
  {Wiringa}}, \bibinfo {author} {\bibfnamefont {R.~A.}\ \bibnamefont {Smith}},
  \ and\ \bibinfo {author} {\bibfnamefont {T.~L.}\ \bibnamefont {Ainsworth}},\
  }\href@noop {} {\bibfield  {journal} {\bibinfo  {journal} {Phys. Rev. C}\
  }\textbf {\bibinfo {volume} {29}},\ \bibinfo {pages} {1207} (\bibinfo {year}
  {1984})}\BibitemShut {NoStop}%
\bibitem [{\citenamefont {Feenberg}(1969)}]{FeenbergBook}%
  \BibitemOpen
  \bibfield  {author} {\bibinfo {author} {\bibfnamefont {E.}~\bibnamefont
  {Feenberg}},\ }\href@noop {} {\emph {\bibinfo {title} {Theory of {Q}uantum
  Fluids}}}\ (\bibinfo  {publisher} {Academic},\ \bibinfo {address} {New
  York},\ \bibinfo {year} {1969})\BibitemShut {NoStop}%
\bibitem [{\citenamefont {Kalos}\ \emph {et~al.}(1974)\citenamefont {Kalos},
  \citenamefont {Levesque},\ and\ \citenamefont {Verlet}}]{KalosLevVer}%
  \BibitemOpen
  \bibfield  {author} {\bibinfo {author} {\bibfnamefont {M.}~\bibnamefont
  {Kalos}}, \bibinfo {author} {\bibfnamefont {D.}~\bibnamefont {Levesque}}, \
  and\ \bibinfo {author} {\bibfnamefont {L.}~\bibnamefont {Verlet}},\
  }\href@noop {} {\bibfield  {journal} {\bibinfo  {journal} {Phys. Rev. A}\
  }\textbf {\bibinfo {volume} {9}},\ \bibinfo {pages} {2178} (\bibinfo {year}
  {1974})}\BibitemShut {NoStop}%
\bibitem [{\citenamefont {Ceperley}(1978)}]{CeperleyVMC}%
  \BibitemOpen
  \bibfield  {author} {\bibinfo {author} {\bibfnamefont {D.~M.}\ \bibnamefont
  {Ceperley}},\ }\href@noop {} {\bibfield  {journal} {\bibinfo  {journal}
  {Phys. Rev. B}\ }\textbf {\bibinfo {volume} {18}},\ \bibinfo {pages} {3126}
  (\bibinfo {year} {1978})}\BibitemShut {NoStop}%
\bibitem [{\citenamefont {Ceperley}(1995)}]{CeperleyRMP}%
  \BibitemOpen
  \bibfield  {author} {\bibinfo {author} {\bibfnamefont {D.~M.}\ \bibnamefont
  {Ceperley}},\ }\href@noop {} {\bibfield  {journal} {\bibinfo  {journal} {Rev.
  Mod. Phys.}\ }\textbf {\bibinfo {volume} {67}},\ \bibinfo {pages} {279}
  (\bibinfo {year} {1995})}\BibitemShut {NoStop}%
\bibitem [{\citenamefont {Boronat}(2002)}]{JordiQFSBook}%
  \BibitemOpen
  \bibfield  {author} {\bibinfo {author} {\bibfnamefont {J.}~\bibnamefont
  {Boronat}},\ }in\ \href@noop {} {\emph {\bibinfo {booktitle} {Microscopic
  Approaches to {Q}uantum Liquids in Confined Geometries}}},\ \bibinfo {editor}
  {edited by\ \bibinfo {editor} {\bibfnamefont {E.}~\bibnamefont {Krotscheck}}\
  and\ \bibinfo {editor} {\bibfnamefont {J.}~\bibnamefont {Navarro}}}\
  (\bibinfo  {publisher} {World Scientific},\ \bibinfo {address} {Singapore},\
  \bibinfo {year} {2002})\ pp.\ \bibinfo {pages} {21--90}\BibitemShut {NoStop}%
\bibitem [{\citenamefont {Morita}(1958)}]{Morita58}%
  \BibitemOpen
  \bibfield  {author} {\bibinfo {author} {\bibfnamefont {T.}~\bibnamefont
  {Morita}},\ }\href@noop {} {\bibfield  {journal} {\bibinfo  {journal} {Progr.
  Theor. Phys.}\ }\textbf {\bibinfo {volume} {20}},\ \bibinfo {pages} {920}
  (\bibinfo {year} {1958})}\BibitemShut {NoStop}%
\bibitem [{\citenamefont {van Leeuwen}\ \emph {et~al.}(1959)\citenamefont {van
  Leeuwen}, \citenamefont {Groeneveld},\ and\ \citenamefont {Boer}}]{LGB}%
  \BibitemOpen
  \bibfield  {author} {\bibinfo {author} {\bibfnamefont {J.~M.~J.}\
  \bibnamefont {van Leeuwen}}, \bibinfo {author} {\bibfnamefont
  {J.}~\bibnamefont {Groeneveld}}, \ and\ \bibinfo {author} {\bibfnamefont
  {J.~D.}\ \bibnamefont {Boer}},\ }\href@noop {} {\bibfield  {journal}
  {\bibinfo  {journal} {Physica}\ }\textbf {\bibinfo {volume} {25}},\ \bibinfo
  {pages} {792} (\bibinfo {year} {1959})}\BibitemShut {NoStop}%
\bibitem [{\citenamefont {Krotscheck}\ and\ \citenamefont
  {Ristig}(1974)}]{Mistig}%
  \BibitemOpen
  \bibfield  {author} {\bibinfo {author} {\bibfnamefont {E.}~\bibnamefont
  {Krotscheck}}\ and\ \bibinfo {author} {\bibfnamefont {M.~L.}\ \bibnamefont
  {Ristig}},\ }\href@noop {} {\bibfield  {journal} {\bibinfo  {journal} {Phys.
  Lett. A}\ }\textbf {\bibinfo {volume} {48}},\ \bibinfo {pages} {17} (\bibinfo
  {year} {1974})}\BibitemShut {NoStop}%
\bibitem [{\citenamefont {Fantoni}\ and\ \citenamefont
  {Rosati}(1975)}]{Fantoni}%
  \BibitemOpen
  \bibfield  {author} {\bibinfo {author} {\bibfnamefont {S.}~\bibnamefont
  {Fantoni}}\ and\ \bibinfo {author} {\bibfnamefont {S.}~\bibnamefont
  {Rosati}},\ }\href@noop {} {\bibfield  {journal} {\bibinfo  {journal} {Nuovo
  Cimento}\ }\textbf {\bibinfo {volume} {25A}},\ \bibinfo {pages} {593}
  (\bibinfo {year} {1975})}\BibitemShut {NoStop}%
\bibitem [{\citenamefont {Jackson}\ \emph {et~al.}(1982)\citenamefont
  {Jackson}, \citenamefont {Lande},\ and\ \citenamefont {Smith}}]{parquet1}%
  \BibitemOpen
  \bibfield  {author} {\bibinfo {author} {\bibfnamefont {A.~D.}\ \bibnamefont
  {Jackson}}, \bibinfo {author} {\bibfnamefont {A.}~\bibnamefont {Lande}}, \
  and\ \bibinfo {author} {\bibfnamefont {R.~A.}\ \bibnamefont {Smith}},\
  }\href@noop {} {\bibfield  {journal} {\bibinfo  {journal} {Physics Reports}\
  }\textbf {\bibinfo {volume} {86}},\ \bibinfo {pages} {55} (\bibinfo {year}
  {1982})}\BibitemShut {NoStop}%
\bibitem [{\citenamefont {Jackson}\ \emph
  {et~al.}(1985{\natexlab{a}})\citenamefont {Jackson}, \citenamefont {Lande},\
  and\ \citenamefont {Smith}}]{parquet2}%
  \BibitemOpen
  \bibfield  {author} {\bibinfo {author} {\bibfnamefont {A.~D.}\ \bibnamefont
  {Jackson}}, \bibinfo {author} {\bibfnamefont {A.}~\bibnamefont {Lande}}, \
  and\ \bibinfo {author} {\bibfnamefont {R.~A.}\ \bibnamefont {Smith}},\
  }\href@noop {} {\bibfield  {journal} {\bibinfo  {journal} {Phys. Rev. Lett.}\
  }\textbf {\bibinfo {volume} {54}},\ \bibinfo {pages} {1469} (\bibinfo {year}
  {1985}{\natexlab{a}})}\BibitemShut {NoStop}%
\bibitem [{\citenamefont {Krotscheck}\ \emph {et~al.}(1986)\citenamefont
  {Krotscheck}, \citenamefont {Smith},\ and\ \citenamefont
  {Jackson}}]{parquet3}%
  \BibitemOpen
  \bibfield  {author} {\bibinfo {author} {\bibfnamefont {E.}~\bibnamefont
  {Krotscheck}}, \bibinfo {author} {\bibfnamefont {R.~A.}\ \bibnamefont
  {Smith}}, \ and\ \bibinfo {author} {\bibfnamefont {A.~D.}\ \bibnamefont
  {Jackson}},\ }\href@noop {} {\bibfield  {journal} {\bibinfo  {journal} {Phys.
  Rev. A}\ }\textbf {\bibinfo {volume} {33}},\ \bibinfo {pages} {3535}
  (\bibinfo {year} {1986})}\BibitemShut {NoStop}%
\bibitem [{\citenamefont {Fantoni}\ and\ \citenamefont
  {Rosati}(1977)}]{FantoniSpins}%
  \BibitemOpen
  \bibfield  {author} {\bibinfo {author} {\bibfnamefont {S.}~\bibnamefont
  {Fantoni}}\ and\ \bibinfo {author} {\bibfnamefont {S.}~\bibnamefont
  {Rosati}},\ }\href@noop {} {\bibfield  {journal} {\bibinfo  {journal} {Nuovo
  Cimento}\ }\textbf {\bibinfo {volume} {43A}},\ \bibinfo {pages} {413}
  (\bibinfo {year} {1977})}\BibitemShut {NoStop}%
\bibitem [{\citenamefont {Pandharipande}\ and\ \citenamefont
  {Wiringa}(1979{\natexlab{a}})}]{IndianSpins}%
  \BibitemOpen
  \bibfield  {author} {\bibinfo {author} {\bibfnamefont {V.~R.}\ \bibnamefont
  {Pandharipande}}\ and\ \bibinfo {author} {\bibfnamefont {R.~B.}\ \bibnamefont
  {Wiringa}},\ }\href@noop {} {\bibfield  {journal} {\bibinfo  {journal} {Rev.
  Mod. Phys.}\ }\textbf {\bibinfo {volume} {51}},\ \bibinfo {pages} {821}
  (\bibinfo {year} {1979}{\natexlab{a}})}\BibitemShut {NoStop}%
\bibitem [{\citenamefont {Wiringa}\ and\ \citenamefont
  {Pandharipande}(1978)}]{Wiri78}%
  \BibitemOpen
  \bibfield  {author} {\bibinfo {author} {\bibfnamefont {R.~B.}\ \bibnamefont
  {Wiringa}}\ and\ \bibinfo {author} {\bibfnamefont {V.~R.}\ \bibnamefont
  {Pandharipande}},\ }\href@noop {} {\bibfield  {journal} {\bibinfo  {journal}
  {Nucl. Phys. A}\ }\textbf {\bibinfo {volume} {299}},\ \bibinfo {pages} {1}
  (\bibinfo {year} {1978})}\BibitemShut {NoStop}%
\bibitem [{\citenamefont {Krotscheck}(1988)}]{SpinTwist}%
  \BibitemOpen
  \bibfield  {author} {\bibinfo {author} {\bibfnamefont {E.}~\bibnamefont
  {Krotscheck}},\ }\href@noop {} {\bibfield  {journal} {\bibinfo  {journal}
  {Nucl. Phys. A}\ }\textbf {\bibinfo {volume} {482}},\ \bibinfo {pages} {617}
  (\bibinfo {year} {1988})}\BibitemShut {NoStop}%
\bibitem [{\citenamefont {Scott}\ and\ \citenamefont
  {Moszkowski}(1962)}]{ScottMozowski}%
  \BibitemOpen
  \bibfield  {author} {\bibinfo {author} {\bibfnamefont {B.~L.}\ \bibnamefont
  {Scott}}\ and\ \bibinfo {author} {\bibfnamefont {S.~A.}\ \bibnamefont
  {Moszkowski}},\ }\href@noop {} {\bibfield  {journal} {\bibinfo  {journal}
  {Nucl. Phys.}\ }\textbf {\bibinfo {volume} {29}},\ \bibinfo {pages} {665}
  (\bibinfo {year} {1962})}\BibitemShut {NoStop}%
\bibitem [{\citenamefont {Pandharipande}\ and\ \citenamefont
  {Bethe}(1973)}]{PandharipandeBethe}%
  \BibitemOpen
  \bibfield  {author} {\bibinfo {author} {\bibfnamefont {V.~R.}\ \bibnamefont
  {Pandharipande}}\ and\ \bibinfo {author} {\bibfnamefont {H.~A.}\ \bibnamefont
  {Bethe}},\ }\href@noop {} {\bibfield  {journal} {\bibinfo  {journal} {Phys.
  Rev. C}\ }\textbf {\bibinfo {volume} {7}},\ \bibinfo {pages} {1312} (\bibinfo
  {year} {1973})}\BibitemShut {NoStop}%
\bibitem [{\citenamefont {Pandharipande}\ and\ \citenamefont
  {Wiringa}(1979{\natexlab{b}})}]{PAW79}%
  \BibitemOpen
  \bibfield  {author} {\bibinfo {author} {\bibfnamefont {V.~R.}\ \bibnamefont
  {Pandharipande}}\ and\ \bibinfo {author} {\bibfnamefont {R.~B.}\ \bibnamefont
  {Wiringa}},\ }\href@noop {} {\bibfield  {journal} {\bibinfo  {journal} {Rev.
  Mod. Phys.}\ }\textbf {\bibinfo {volume} {51}},\ \bibinfo {pages} {821}
  (\bibinfo {year} {1979}{\natexlab{b}})}\BibitemShut {NoStop}%
\bibitem [{\citenamefont {Smith}\ and\ \citenamefont
  {Jackson}(1988)}]{SmithSpin}%
  \BibitemOpen
  \bibfield  {author} {\bibinfo {author} {\bibfnamefont {R.~A.}\ \bibnamefont
  {Smith}}\ and\ \bibinfo {author} {\bibfnamefont {A.~D.}\ \bibnamefont
  {Jackson}},\ }\href@noop {} {\bibfield  {journal} {\bibinfo  {journal} {Nucl.
  Phys. A}\ }\textbf {\bibinfo {volume} {476}},\ \bibinfo {pages} {448}
  (\bibinfo {year} {1988})}\BibitemShut {NoStop}%
\bibitem [{\citenamefont {Baldo}\ \emph {et~al.}(2012)\citenamefont {Baldo},
  \citenamefont {Polls}, \citenamefont {Rios}, \citenamefont {Schulze},\ and\
  \citenamefont {Vida{\~n}a}}]{Baldo2012}%
  \BibitemOpen
  \bibfield  {author} {\bibinfo {author} {\bibfnamefont {M.}~\bibnamefont
  {Baldo}}, \bibinfo {author} {\bibfnamefont {A.}~\bibnamefont {Polls}},
  \bibinfo {author} {\bibfnamefont {A.}~\bibnamefont {Rios}}, \bibinfo {author}
  {\bibfnamefont {H.-J.}\ \bibnamefont {Schulze}}, \ and\ \bibinfo {author}
  {\bibfnamefont {I.}~\bibnamefont {Vida{\~n}a}},\ }\href@noop {} {\bibfield
  {journal} {\bibinfo  {journal} {Phys. Rev. C}\ }\textbf {\bibinfo {volume}
  {86}},\ \bibinfo {pages} {064001} (\bibinfo {year} {2012})}\BibitemShut
  {NoStop}%
\bibitem [{\citenamefont {Bishop}(1995)}]{BishopValencia}%
  \BibitemOpen
  \bibfield  {author} {\bibinfo {author} {\bibfnamefont {R.~F.}\ \bibnamefont
  {Bishop}},\ }in\ \href@noop {} {\emph {\bibinfo {booktitle} {Condensed Matter
  Theories}}},\ Vol.~\bibinfo {volume} {10},\ \bibinfo {editor} {edited by\
  \bibinfo {editor} {\bibfnamefont {M.}~\bibnamefont {Casas}}, \bibinfo
  {editor} {\bibfnamefont {J.}~\bibnamefont {Navarro}}, \ and\ \bibinfo
  {editor} {\bibfnamefont {A.}~\bibnamefont {Polls}}}\ (\bibinfo  {publisher}
  {Nova Science Publishers},\ \bibinfo {address} {Commack, New York},\ \bibinfo
  {year} {1995})\ pp.\ \bibinfo {pages} {483--508}\BibitemShut {NoStop}%
\bibitem [{\citenamefont {Krotscheck}(1994)}]{PairDFT}%
  \BibitemOpen
  \bibfield  {author} {\bibinfo {author} {\bibfnamefont {E.}~\bibnamefont
  {Krotscheck}},\ }\href@noop {} {\bibfield  {journal} {\bibinfo  {journal}
  {Phys. Lett. A}\ }\textbf {\bibinfo {volume} {190}},\ \bibinfo {pages} {201}
  (\bibinfo {year} {1994})}\BibitemShut {NoStop}%
\bibitem [{\citenamefont {Fan}\ and\ \citenamefont
  {Krotscheck}(2019)}]{fullbcs}%
  \BibitemOpen
  \bibfield  {author} {\bibinfo {author} {\bibfnamefont {H.-H.}\ \bibnamefont
  {Fan}}\ and\ \bibinfo {author} {\bibfnamefont {E.}~\bibnamefont
  {Krotscheck}},\ }\href@noop {} {\bibfield  {journal} {\bibinfo  {journal}
  {Physics Reports}\ }\textbf {\bibinfo {volume} {823}},\ \bibinfo {pages} {1}
  (\bibinfo {year} {2019})}\BibitemShut {NoStop}%
\bibitem [{\citenamefont {Krotscheck}(1986)}]{EKthree}%
  \BibitemOpen
  \bibfield  {author} {\bibinfo {author} {\bibfnamefont {E.}~\bibnamefont
  {Krotscheck}},\ }\href@noop {} {\bibfield  {journal} {\bibinfo  {journal}
  {Phys. Rev. B}\ }\textbf {\bibinfo {volume} {33}},\ \bibinfo {pages} {3158}
  (\bibinfo {year} {1986})}\BibitemShut {NoStop}%
\bibitem [{\citenamefont {Jackson}\ \emph
  {et~al.}(1985{\natexlab{b}})\citenamefont {Jackson}, \citenamefont {Lande},
  \citenamefont {Guitink},\ and\ \citenamefont {Smith}}]{TripletParquet}%
  \BibitemOpen
  \bibfield  {author} {\bibinfo {author} {\bibfnamefont {A.~D.}\ \bibnamefont
  {Jackson}}, \bibinfo {author} {\bibfnamefont {A.}~\bibnamefont {Lande}},
  \bibinfo {author} {\bibfnamefont {R.~W.}\ \bibnamefont {Guitink}}, \ and\
  \bibinfo {author} {\bibfnamefont {R.~A.}\ \bibnamefont {Smith}},\ }\href@noop
  {} {\bibfield  {journal} {\bibinfo  {journal} {Phys. Rev. B}\ }\textbf
  {\bibinfo {volume} {31}},\ \bibinfo {pages} {403} (\bibinfo {year}
  {1985}{\natexlab{b}})}\BibitemShut {NoStop}%
\bibitem [{\citenamefont {Lantto}\ and\ \citenamefont
  {Siemens}(1977)}]{LanttoSiemens}%
  \BibitemOpen
  \bibfield  {author} {\bibinfo {author} {\bibfnamefont {L.~J.}\ \bibnamefont
  {Lantto}}\ and\ \bibinfo {author} {\bibfnamefont {P.~J.}\ \bibnamefont
  {Siemens}},\ }\href@noop {} {\bibfield  {journal} {\bibinfo  {journal} {Phys.
  Lett. B}\ }\textbf {\bibinfo {volume} {68}},\ \bibinfo {pages} {308}
  (\bibinfo {year} {1977})}\BibitemShut {NoStop}%
\bibitem [{\citenamefont {Sim}\ \emph {et~al.}(1970)\citenamefont {Sim},
  \citenamefont {Woo},\ and\ \citenamefont {Buchler}}]{Woo70}%
  \BibitemOpen
  \bibfield  {author} {\bibinfo {author} {\bibfnamefont {H.~K.}\ \bibnamefont
  {Sim}}, \bibinfo {author} {\bibfnamefont {C.-W.}\ \bibnamefont {Woo}}, \ and\
  \bibinfo {author} {\bibfnamefont {J.~R.}\ \bibnamefont {Buchler}},\
  }\href@noop {} {\bibfield  {journal} {\bibinfo  {journal} {Phys. Rev. A}\
  }\textbf {\bibinfo {volume} {2}},\ \bibinfo {pages} {2024} (\bibinfo {year}
  {1970})}\BibitemShut {NoStop}%
\bibitem [{\citenamefont {Clark}(1979)}]{Johnreview}%
  \BibitemOpen
  \bibfield  {author} {\bibinfo {author} {\bibfnamefont {J.~W.}\ \bibnamefont
  {Clark}},\ }in\ \href@noop {} {\emph {\bibinfo {booktitle} {Progress in
  Particle and Nuclear Physics}}},\ Vol.~\bibinfo {volume} {2},\ \bibinfo
  {editor} {edited by\ \bibinfo {editor} {\bibfnamefont {D.~H.}\ \bibnamefont
  {Wilkinson}}}\ (\bibinfo  {publisher} {Pergamon Press Ltd.},\ \bibinfo
  {address} {Oxford},\ \bibinfo {year} {1979})\ pp.\ \bibinfo {pages}
  {89--199}\BibitemShut {NoStop}%
\bibitem [{\citenamefont {Krotscheck}(2000)}]{polish}%
  \BibitemOpen
  \bibfield  {author} {\bibinfo {author} {\bibfnamefont {E.}~\bibnamefont
  {Krotscheck}},\ }\href@noop {} {\bibfield  {journal} {\bibinfo  {journal} {J.
  Low Temp. Phys.}\ }\textbf {\bibinfo {volume} {119}},\ \bibinfo {pages} {103}
  (\bibinfo {year} {2000})}\BibitemShut {NoStop}%
\bibitem [{\citenamefont {Krotscheck}\ and\ \citenamefont
  {Saarela}(1993)}]{mixmonster}%
  \BibitemOpen
  \bibfield  {author} {\bibinfo {author} {\bibfnamefont {E.}~\bibnamefont
  {Krotscheck}}\ and\ \bibinfo {author} {\bibfnamefont {M.}~\bibnamefont
  {Saarela}},\ }\href@noop {} {\bibfield  {journal} {\bibinfo  {journal}
  {Physics Reports}\ }\textbf {\bibinfo {volume} {232}},\ \bibinfo {pages} {1}
  (\bibinfo {year} {1993})}\BibitemShut {NoStop}%
\bibitem [{\citenamefont {Owen}(1981)}]{OwenVar}%
  \BibitemOpen
  \bibfield  {author} {\bibinfo {author} {\bibfnamefont {J.~C.}\ \bibnamefont
  {Owen}},\ }\href@noop {} {\bibfield  {journal} {\bibinfo  {journal} {Phys.
  Rev. B}\ }\textbf {\bibinfo {volume} {23}},\ \bibinfo {pages} {2169}
  (\bibinfo {year} {1981})}\BibitemShut {NoStop}%
\bibitem [{\citenamefont {Fetter}\ and\ \citenamefont
  {Walecka}(1971)}]{FetterWalecka}%
  \BibitemOpen
  \bibfield  {author} {\bibinfo {author} {\bibfnamefont {A.~L.}\ \bibnamefont
  {Fetter}}\ and\ \bibinfo {author} {\bibfnamefont {J.~D.}\ \bibnamefont
  {Walecka}},\ }\href@noop {} {\emph {\bibinfo {title} {{Q}uantum Theory of
  Many-Particle Systems}}}\ (\bibinfo  {publisher} {McGraw-Hill},\ \bibinfo
  {address} {New York},\ \bibinfo {year} {1971})\BibitemShut {NoStop}%
\bibitem [{\citenamefont {Bethe}\ and\ \citenamefont
  {Goldstone}(1957)}]{BetheGoldstone57}%
  \BibitemOpen
  \bibfield  {author} {\bibinfo {author} {\bibfnamefont {H.~A.}\ \bibnamefont
  {Bethe}}\ and\ \bibinfo {author} {\bibfnamefont {J.}~\bibnamefont
  {Goldstone}},\ }\href@noop {} {\bibfield  {journal} {\bibinfo  {journal}
  {Proc. R. Soc. London, Ser. A}\ }\textbf {\bibinfo {volume} {238}},\ \bibinfo
  {pages} {551} (\bibinfo {year} {1957})}\BibitemShut {NoStop}%
\bibitem [{\citenamefont {Ripka}(1979)}]{Rip79}%
  \BibitemOpen
  \bibfield  {author} {\bibinfo {author} {\bibfnamefont {G.}~\bibnamefont
  {Ripka}},\ }\href@noop {} {\bibfield  {journal} {\bibinfo  {journal} {Nucl.
  Phys. A}\ }\textbf {\bibinfo {volume} {314}},\ \bibinfo {pages} {115}
  (\bibinfo {year} {1979})}\BibitemShut {NoStop}%
\bibitem [{\citenamefont {Friman}\ \emph {et~al.}(1982)\citenamefont {Friman},
  \citenamefont {Niskanen},\ and\ \citenamefont {Nyman}}]{FNN82}%
  \BibitemOpen
  \bibfield  {author} {\bibinfo {author} {\bibfnamefont {B.~L.}\ \bibnamefont
  {Friman}}, \bibinfo {author} {\bibfnamefont {J.}~\bibnamefont {Niskanen}}, \
  and\ \bibinfo {author} {\bibfnamefont {E.~M.}\ \bibnamefont {Nyman}},\
  }\href@noop {} {\bibfield  {journal} {\bibinfo  {journal} {Nucl. Phys. A}\
  }\textbf {\bibinfo {volume} {383}},\ \bibinfo {pages} {285} (\bibinfo {year}
  {1982})}\BibitemShut {NoStop}%
\bibitem [{\citenamefont {Thouless}(1972)}]{ThoulessBook}%
  \BibitemOpen
  \bibfield  {author} {\bibinfo {author} {\bibfnamefont {D.~J.}\ \bibnamefont
  {Thouless}},\ }\href@noop {} {\emph {\bibinfo {title} {The quantum mechanics
  of many-body systems}}},\ \bibinfo {edition} {2nd}\ ed.\ (\bibinfo
  {publisher} {Academic Press},\ \bibinfo {address} {New York},\ \bibinfo
  {year} {1972})\BibitemShut {NoStop}%
\bibitem [{\citenamefont {Gezerlis}\ \emph {et~al.}(2014)\citenamefont
  {Gezerlis}, \citenamefont {Pethick},\ and\ \citenamefont
  {Schwenk}}]{Gezerlis2014}%
  \BibitemOpen
  \bibfield  {author} {\bibinfo {author} {\bibfnamefont {A.}~\bibnamefont
  {Gezerlis}}, \bibinfo {author} {\bibfnamefont {C.~J.}\ \bibnamefont
  {Pethick}}, \ and\ \bibinfo {author} {\bibfnamefont {A.}~\bibnamefont
  {Schwenk}},\ }in\ \href@noop {} {\emph {\bibinfo {booktitle} {Novel
  Superfluids}}},\ Vol.~\bibinfo {volume} {2},\ \bibinfo {editor} {edited by\
  \bibinfo {editor} {\bibfnamefont {K.~H.}\ \bibnamefont {Bennemann}}\ and\
  \bibinfo {editor} {\bibfnamefont {J.~B.}\ \bibnamefont {Ketterson}}}\
  (\bibinfo  {publisher} {Oxford University Press},\ \bibinfo {year} {2014})\
  Chap.~\bibinfo {chapter} {22}, pp.\ \bibinfo {pages} {580--615}\BibitemShut
  {NoStop}%
\bibitem [{\citenamefont {Gezerlis}\ and\ \citenamefont
  {Carlson}(2008)}]{GC2008}%
  \BibitemOpen
  \bibfield  {author} {\bibinfo {author} {\bibfnamefont {A.}~\bibnamefont
  {Gezerlis}}\ and\ \bibinfo {author} {\bibfnamefont {J.}~\bibnamefont
  {Carlson}},\ }\href@noop {} {\bibfield  {journal} {\bibinfo  {journal} {Phys.
  Rev. C}\ }\textbf {\bibinfo {volume} {77}},\ \bibinfo {pages} {032801}
  (\bibinfo {year} {2008})}\BibitemShut {NoStop}%
\bibitem [{\citenamefont {Fan}\ \emph {et~al.}(2017)\citenamefont {Fan},
  \citenamefont {Krotscheck},\ and\ \citenamefont {Clark}}]{ectpaper}%
  \BibitemOpen
  \bibfield  {author} {\bibinfo {author} {\bibfnamefont {H.-H.}\ \bibnamefont
  {Fan}}, \bibinfo {author} {\bibfnamefont {E.}~\bibnamefont {Krotscheck}}, \
  and\ \bibinfo {author} {\bibfnamefont {J.~W.}\ \bibnamefont {Clark}},\
  }\href@noop {} {\bibfield  {journal} {\bibinfo  {journal} {J. Low Temp.
  Phys.}\ }\textbf {\bibinfo {volume} {189}},\ \bibinfo {pages} {470} (\bibinfo
  {year} {2017})}\BibitemShut {NoStop}%
\bibitem [{\citenamefont {Gandolfi}\ \emph {et~al.}(2009)\citenamefont
  {Gandolfi}, \citenamefont {Illarionov}, \citenamefont {Schmidt},
  \citenamefont {Pederiva},\ and\ \citenamefont {Fantoni}}]{Gandolfi2009a}%
  \BibitemOpen
  \bibfield  {author} {\bibinfo {author} {\bibfnamefont {S.}~\bibnamefont
  {Gandolfi}}, \bibinfo {author} {\bibfnamefont {A.~Y.}\ \bibnamefont
  {Illarionov}}, \bibinfo {author} {\bibfnamefont {K.~E.}\ \bibnamefont
  {Schmidt}}, \bibinfo {author} {\bibfnamefont {F.}~\bibnamefont {Pederiva}}, \
  and\ \bibinfo {author} {\bibfnamefont {S.}~\bibnamefont {Fantoni}},\
  }\href@noop {} {\bibfield  {journal} {\bibinfo  {journal} {Phys. Rev. C}\
  }\textbf {\bibinfo {volume} {79}},\ \bibinfo {pages} {054005} (\bibinfo
  {year} {2009})}\BibitemShut {NoStop}%
\bibitem [{\citenamefont {Gonz\'alez~Trotter}\ \emph
  {et~al.}(1999)\citenamefont {Gonz\'alez~Trotter}, \citenamefont {Salinas},
  \citenamefont {Chen}, \citenamefont {Crowell}, \citenamefont {Gl\"ockle},
  \citenamefont {Howell}, \citenamefont {Roper}, \citenamefont {Schmidt},
  \citenamefont {\ifmmode~\check{S}\else \v{S}\fi{}laus}, \citenamefont {Tang},
  \citenamefont {Tornow}, \citenamefont {Walter}, \citenamefont {Wita\l{}a},\
  and\ \citenamefont {Zhou}}]{PhysRevLett.83.3788}%
  \BibitemOpen
  \bibfield  {author} {\bibinfo {author} {\bibfnamefont {D.~E.}\ \bibnamefont
  {Gonz\'alez~Trotter}}, \bibinfo {author} {\bibfnamefont {F.}~\bibnamefont
  {Salinas}}, \bibinfo {author} {\bibfnamefont {Q.}~\bibnamefont {Chen}},
  \bibinfo {author} {\bibfnamefont {A.~S.}\ \bibnamefont {Crowell}}, \bibinfo
  {author} {\bibfnamefont {W.}~\bibnamefont {Gl\"ockle}}, \bibinfo {author}
  {\bibfnamefont {C.~R.}\ \bibnamefont {Howell}}, \bibinfo {author}
  {\bibfnamefont {C.~D.}\ \bibnamefont {Roper}}, \bibinfo {author}
  {\bibfnamefont {D.}~\bibnamefont {Schmidt}}, \bibinfo {author} {\bibfnamefont
  {I.}~\bibnamefont {\ifmmode~\check{S}\else \v{S}\fi{}laus}}, \bibinfo
  {author} {\bibfnamefont {H.}~\bibnamefont {Tang}}, \bibinfo {author}
  {\bibfnamefont {W.}~\bibnamefont {Tornow}}, \bibinfo {author} {\bibfnamefont
  {R.~L.}\ \bibnamefont {Walter}}, \bibinfo {author} {\bibfnamefont
  {H.}~\bibnamefont {Wita\l{}a}}, \ and\ \bibinfo {author} {\bibfnamefont
  {Z.}~\bibnamefont {Zhou}},\ }\href {\doibase 10.1103/PhysRevLett.83.3788}
  {\bibfield  {journal} {\bibinfo  {journal} {Phys. Rev. Lett.}\ }\textbf
  {\bibinfo {volume} {83}},\ \bibinfo {pages} {3788} (\bibinfo {year}
  {1999})}\BibitemShut {NoStop}%
\bibitem [{\citenamefont {Bethe}\ \emph {et~al.}(1963)\citenamefont {Bethe},
  \citenamefont {Brandow},\ and\ \citenamefont
  {Petschek}}]{BetheBrandowPetschek}%
  \BibitemOpen
  \bibfield  {author} {\bibinfo {author} {\bibfnamefont {H.~A.}\ \bibnamefont
  {Bethe}}, \bibinfo {author} {\bibfnamefont {B.~H.}\ \bibnamefont {Brandow}},
  \ and\ \bibinfo {author} {\bibfnamefont {A.~G.}\ \bibnamefont {Petschek}},\
  }\href@noop {} {\bibfield  {journal} {\bibinfo  {journal} {Phys. Rev.}\
  }\textbf {\bibinfo {volume} {129}},\ \bibinfo {pages} {225} (\bibinfo {year}
  {1963})}\BibitemShut {NoStop}%
\bibitem [{\citenamefont {Krotscheck}(1977{\natexlab{a}})}]{Kro77}%
  \BibitemOpen
  \bibfield  {author} {\bibinfo {author} {\bibfnamefont {E.}~\bibnamefont
  {Krotscheck}},\ }\href@noop {} {\bibfield  {journal} {\bibinfo  {journal}
  {Nucl. Phys. A}\ }\textbf {\bibinfo {volume} {293}},\ \bibinfo {pages} {293}
  (\bibinfo {year} {1977}{\natexlab{a}})}\BibitemShut {NoStop}%
\bibitem [{\citenamefont {Wambach}\ \emph {et~al.}(1993)\citenamefont
  {Wambach}, \citenamefont {Ainsworth},\ and\ \citenamefont {Pines}}]{Wam93}%
  \BibitemOpen
  \bibfield  {author} {\bibinfo {author} {\bibfnamefont {J.}~\bibnamefont
  {Wambach}}, \bibinfo {author} {\bibfnamefont {T.}~\bibnamefont {Ainsworth}},
  \ and\ \bibinfo {author} {\bibfnamefont {D.}~\bibnamefont {Pines}},\
  }\href@noop {} {\bibfield  {journal} {\bibinfo  {journal} {Nucl. Phys. A}\
  }\textbf {\bibinfo {volume} {555}},\ \bibinfo {pages} {128} (\bibinfo {year}
  {1993})}\BibitemShut {NoStop}%
\bibitem [{\citenamefont {Sedrakian}\ and\ \citenamefont
  {Clark}(2018)}]{SedrakianClarkBCSReview}%
  \BibitemOpen
  \bibfield  {author} {\bibinfo {author} {\bibfnamefont {A.}~\bibnamefont
  {Sedrakian}}\ and\ \bibinfo {author} {\bibfnamefont {J.~W.}\ \bibnamefont
  {Clark}},\ }\href@noop {} {\enquote {\bibinfo {title} {Superfluidity in
  nuclear systems and neutron stars},}\ } (\bibinfo {year} {2018}),\ \bibinfo
  {note} {arXiv:1802.00017}\BibitemShut {NoStop}%
\bibitem [{\citenamefont {Krotscheck}(1977{\natexlab{b}})}]{EKVar}%
  \BibitemOpen
  \bibfield  {author} {\bibinfo {author} {\bibfnamefont {E.}~\bibnamefont
  {Krotscheck}},\ }\href@noop {} {\bibfield  {journal} {\bibinfo  {journal}
  {Phys. Rev. A}\ }\textbf {\bibinfo {volume} {15}},\ \bibinfo {pages} {397}
  (\bibinfo {year} {1977}{\natexlab{b}})}\BibitemShut {NoStop}%
\bibitem [{\citenamefont {Jackson}\ and\ \citenamefont
  {Smith}(1987)}]{parquet5}%
  \BibitemOpen
  \bibfield  {author} {\bibinfo {author} {\bibfnamefont {A.~D.}\ \bibnamefont
  {Jackson}}\ and\ \bibinfo {author} {\bibfnamefont {R.~A.}\ \bibnamefont
  {Smith}},\ }\href@noop {} {\bibfield  {journal} {\bibinfo  {journal} {Phys.
  Rev. A}\ }\textbf {\bibinfo {volume} {36}},\ \bibinfo {pages} {2517}
  (\bibinfo {year} {1987})}\BibitemShut {NoStop}%
\bibitem [{\citenamefont {Campbell}\ \emph {et~al.}(1996)\citenamefont
  {Campbell}, \citenamefont {Folk},\ and\ \citenamefont
  {Krotscheck}}]{lowdens}%
  \BibitemOpen
  \bibfield  {author} {\bibinfo {author} {\bibfnamefont {C.~E.}\ \bibnamefont
  {Campbell}}, \bibinfo {author} {\bibfnamefont {R.}~\bibnamefont {Folk}}, \
  and\ \bibinfo {author} {\bibfnamefont {E.}~\bibnamefont {Krotscheck}},\
  }\href@noop {} {\bibfield  {journal} {\bibinfo  {journal} {J. Low Temp.
  Phys.}\ }\textbf {\bibinfo {volume} {105}},\ \bibinfo {pages} {13} (\bibinfo
  {year} {1996})}\BibitemShut {NoStop}%
\bibitem [{\citenamefont {Krotscheck}\ and\ \citenamefont
  {Clark}(1980)}]{HNCBCS}%
  \BibitemOpen
  \bibfield  {author} {\bibinfo {author} {\bibfnamefont {E.}~\bibnamefont
  {Krotscheck}}\ and\ \bibinfo {author} {\bibfnamefont {J.~W.}\ \bibnamefont
  {Clark}},\ }\href@noop {} {\bibfield  {journal} {\bibinfo  {journal} {Nucl.
  Phys. A}\ }\textbf {\bibinfo {volume} {333}},\ \bibinfo {pages} {77}
  (\bibinfo {year} {1980})}\BibitemShut {NoStop}%
\bibitem [{\citenamefont {Bardeen}\ \emph {et~al.}(1957)\citenamefont
  {Bardeen}, \citenamefont {Cooper},\ and\ \citenamefont {Schrieffer}}]{bcs}%
  \BibitemOpen
  \bibfield  {author} {\bibinfo {author} {\bibfnamefont {J.}~\bibnamefont
  {Bardeen}}, \bibinfo {author} {\bibfnamefont {L.~N.}\ \bibnamefont {Cooper}},
  \ and\ \bibinfo {author} {\bibfnamefont {J.~R.}\ \bibnamefont {Schrieffer}},\
  }\href@noop {} {\bibfield  {journal} {\bibinfo  {journal} {Phys. Rev.}\
  }\textbf {\bibinfo {volume} {108}},\ \bibinfo {pages} {1175} (\bibinfo {year}
  {1957})}\BibitemShut {NoStop}%
\bibitem [{\citenamefont {Krotscheck}\ \emph {et~al.}(1998)\citenamefont
  {Krotscheck}, \citenamefont {Paaso}, \citenamefont {Saarela}, \citenamefont
  {Sch{\"o}rkhuber},\ and\ \citenamefont {Zillich}}]{mixmass}%
  \BibitemOpen
  \bibfield  {author} {\bibinfo {author} {\bibfnamefont {E.}~\bibnamefont
  {Krotscheck}}, \bibinfo {author} {\bibfnamefont {J.}~\bibnamefont {Paaso}},
  \bibinfo {author} {\bibfnamefont {M.}~\bibnamefont {Saarela}}, \bibinfo
  {author} {\bibfnamefont {K.}~\bibnamefont {Sch{\"o}rkhuber}}, \ and\ \bibinfo
  {author} {\bibfnamefont {R.}~\bibnamefont {Zillich}},\ }\href@noop {}
  {\bibfield  {journal} {\bibinfo  {journal} {Phys. Rev. B}\ }\textbf {\bibinfo
  {volume} {58}},\ \bibinfo {pages} {12282} (\bibinfo {year}
  {1998})}\BibitemShut {NoStop}%
\bibitem [{\citenamefont {Krotscheck}(1982)}]{rings}%
  \BibitemOpen
  \bibfield  {author} {\bibinfo {author} {\bibfnamefont {E.}~\bibnamefont
  {Krotscheck}},\ }\href@noop {} {\bibfield  {journal} {\bibinfo  {journal}
  {Phys. Rev. A}\ }\textbf {\bibinfo {volume} {26}},\ \bibinfo {pages} {3536}
  (\bibinfo {year} {1982})}\BibitemShut {NoStop}%
\end{thebibliography}
%
\end{document}